\documentclass{sig-alternate}

\usepackage{amsmath}
\usepackage{listings}
\usepackage{subfigure}
\usepackage{hyperref}

\hypersetup{pdftitle={Performance Portability Study of Linear Algebra Kernels in OpenCL},
            pdfauthor={Karl Rupp, Philippe Tillet, Florian Rudolf, Josef Weinbub, Tibor Grasser, Ansgar J\"ungel}}

\usepackage{listings}
\usepackage{xcolor}

\definecolor{mygreen}{rgb}{0,0.6,0}
\definecolor{mygray}{rgb}{0.5,0.5,0.5}
\definecolor{mymauve}{rgb}{0.58,0,0.82}
\definecolor{ghost white}{rgb}{0.972549,0.972549,1.000000} 

\begin{document}
%
% --- Author Metadata here ---
\conferenceinfo{IWOCL}{'14, Bristol, UK}
%\CopyrightYear{2007} % Allows default copyright year (20XX) to be over-ridden - IF NEED BE.
%\crdata{0-12345-67-8/90/01}  % Allows default copyright data (0-89791-88-6/97/05) to be over-ridden - IF NEED BE.
% --- End of Author Metadata ---

\title{Performance Portability Study \\ of Linear Algebra Kernels in OpenCL \vspace*{-0.2cm}}

%\subtitle{[Extended Abstract]
%\titlenote{A full version of this paper is available as
%\textit{Author's Guide to Preparing ACM SIG Proceedings Using
%\LaTeX$2_\epsilon$\ and BibTeX} at


\numberofauthors{6} %  in this sample file, there are a *total*
\author{
% 1st. author
\alignauthor
Karl Rupp\\
       \affaddr{Institute for Microelectronics, TU Wien}\\
       \affaddr{Gu\ss hausstr.~ 27-29/E360}\\
       \affaddr{A-1040 Wien, Austria}\\
       \email{rupp@iue.tuwien.ac.at}
% 2nd. author
\alignauthor
Philippe Tillet\\
       \affaddr{Institute for Microelectronics, TU Wien}\\
       \affaddr{Gu\ss hausstr.~ 27-29/E360}\\
       \affaddr{A-1040 Wien, Austria}\\
       \email{phil.tillet@gmail.com}
% 3rd. author
\alignauthor
Florian Rudolf\\
       \affaddr{Institute for Microelectronics, TU Wien}\\
       \affaddr{Gu\ss hausstr.~ 27-29/E360}\\
       \affaddr{A-1040 Wien, Austria}\\
       \email{rudolf@iue.tuwien.ac.at}
\and  % use '\and' if you need 'another row' of author names
% 4th. author
\alignauthor
Josef Weinbub\\
       \affaddr{Institute for Microelectronics, TU Wien}\\
       \affaddr{Gu\ss hausstr.~ 27-29/E360}\\
       \affaddr{A-1040 Wien, Austria}\\
       \email{weinbub@iue.tuwien.ac.at}
% 5th. author
\alignauthor
Tibor Grasser\\
       \affaddr{Institute for Microelectronics, TU Wien}\\
       \affaddr{Gu\ss hausstr.~ 27-29/E360}\\
       \affaddr{A-1040 Wien, Austria}\\
       \email{grasser@iue.tuwien.ac.at}
% 6th. author
\alignauthor Ansgar J\"ungel\\
       \affaddr{Institute for Analysis and Scientific Computing, TU Wien}\\
       \affaddr{Wiedner Hauptstr.~8-10/E101}\\
       \affaddr{A-1040 Wien, Austria}\\
       \email{juengel@asc.tuwien.ac.at}
}
% There's nothing stopping you putting the seventh, eighth, etc.
% author on the opening page (as the 'third row') but we ask,
% for aesthetic reasons that you place these 'additional authors'
% in the \additional authors block, viz.
%\additionalauthors{Additional authors: John Smith (The Th{\o}rv{\"a}ld Group,
%email: {\texttt{jsmith@affiliation.org}}) and Julius P.~Kumquat
%(The Kumquat Consortium, email: {\texttt{jpkumquat@consortium.net}}).}
% Just remember to make sure that the TOTAL number of authors
% is the number that will appear on the first page PLUS the
% number that will appear in the \additionalauthors section.

\maketitle
\begin{abstract}
The performance portability of OpenCL kernel implementations for common memory bandwidth limited linear algebra operations across different hardware generations of the same vendor as well as across vendors is studied.
Certain combinations of kernel implementations and work sizes are found to exhibit good performance across compute kernels, hardware generations, and, to a lesser degree, vendors.
As a consequence, it is demonstrated that the optimization of a single kernel is often sufficient to obtain good performance for a large class of more complicated operations.
\end{abstract}

%A category including the fourth, optional field follows...
%\category{D.2.8}{Software Engineering}{Metrics}[complexity measures, performance measures]

%\terms{OpenCL}


\section{Introduction}
With the introduction of OpenCL \cite{OpenCL}, a unified application programming interface for massively parallel hardware became available, simplifying the implementation of portable software.
However, the portability of code does not automatically imply the portability of performance \cite{McIntosh-Smith:high-performance-drug-screening,Pennycook:OpenCL-performance-portability,Ru:experimental-study-performance-portability-OpenCL,Thoman:automatic-OpenCL-device-characterization,Zhang:Improving-Performance-Portability}, neither across vendors nor within different hardware generations of the same vendor.
To obtain best performance for a given device, autotuning approaches with different degrees of sophistication were proposed, particularly for the optimization of dense matrix-matrix multiplications \cite{Du:CUDA-OpenCL,Tomov2012,Matsumoto2012}.
These approaches perform an exhaustive search in a large space of \emph{kernel configurations}, which we define as the triple consisting of parameterized source code, local work size, and global work size as specified in the OpenCL standard.

Despite the success of autotuning approaches for finding the best kernel configuration for a given device, they also entail excessive execution times. 
While these execution times are often acceptable for software developers and scientific investigations, they represent a considerable burden when exposed to users through software libraries.
An example is the linear algebra library ATLAS \cite{Atlas,Whaley:ATLAS}, which executes the autotuning process during installation on the target machine, eventually taking hours to complete and effectively blocking the machine for that period.
With further increases in the number of cores and more complicated cache hierarchies, such autotuning approaches face severe practical limitations stemming from an explosion of the search space.

This work is motivated by previous work on portable performance for linear algebra operations through a kernel generation approach \cite{Tillet:Performance-portable-linear-algebra} implemented in the free open-source library ViennaCL \cite{Rupp:ViennaCL,ViennaCL}.
Due to the just-in-time compilation capabilities of OpenCL, such a kernel generation can be completely embedded into a software library.
At the same time, the performance is competitive with vendor-tuned libraries on devices from AMD, Intel, and NVIDIA.
The portable performance reported in \cite{Tillet:Performance-portable-linear-algebra} was obtained from in-house autotuning runs for each of the kernels on the respective target hardware.
While this enables optimized kernel configurations for selected hardware, the approach does not scale well to the full range of hardware available on the market:
First, the procedure has to be repeated for each new hardware generation, which is both very time consuming and does not provide a deeper understanding of the underlying hardware.
Second, with the broader support of OpenCL by hardware vendors it becomes practically impossible to repeat the tuning procedure for each device generation available.
To overcome these limitations, it is necessary to better understand the influences of each of the parameters available inside the kernel configurations.

The purpose of this work is to find a strategy for obtaining portable performance across a wide range of possibly unknown target hardware.
To achieve this we systematically study the performance variations obtained for linear algebra kernels simpler than those for the aforementioned matrix-matrix multiplications.
We consider vector and dense matrix-vector operations, from which the basic linear algebra subprograms (BLAS) operations at level 1 and 2 are composed.
More specifically, we consider the vector copy operation ($x \leftarrow y$), a scaled vector addition ($x \leftarrow \alpha y + \beta z$), the inner product of two vectors ($x \leftarrow \langle y, z \rangle$), and the matrix-vector product $x \leftarrow A y$.
Since all operations are limited by the available memory bandwidth, one may expect that it is sufficient to tune just one of these operations and deduce fast implementations for the others.
As will be shown in the remainder of this work, this is indeed possible: The best implementations for the copy operation serve as archetypes for obtaining good performance for the other operations.

To stay within a reasonable scope, no operations related to sparse matrices are considered in this work.
Their optimal data structure depends on the underlying sparsity pattern \cite{baskaran:sparse-matvec-ibm,Bell:SpMV-on-throughput-oriented-processors}, which would add another degree of freedom to the variability of performance due to the underlying hardware and the OpenCL implementation.
Also, only real-valued arithmetic is considered, as OpenCL does not yet define a native type for complex arithmetic and the emulation of complex-valued arithmetic would hence introduce another unnecessary source of variation.

We conduct our study in a hierarchical manner:
First, all kernel configurations for a given operation (vector copy, vector addition, inner product, or dense matrix-vector product) are investigated.
Second, we correlate the performance of kernel configurations across different operations on the same hardware and it is shown that high performance for all kernels can be obtained by running a full optimization for only one of the kernels.
Third, performances across hardware of the same type and vendor, e.g.~GPUs from NVIDIA, are compared, where it is found that good kernel configurations for older hardware often provides good performance on newer devices.
Finally, we compare the performance of kernel configurations across different hardware types and vendors and show that newer hardware generations show better performance portability when compared to older generations.

%The parameterizations we use for the kernel configurations for each of the operations listed above are outlined in Section \ref{sec:parametrization}.
%Section \ref{sec:results} studies the performance obtained for each of the kernel configurations and operations.

\section{Kernel Parameterization} \label{sec:parametrization}
In a massively parallel setting, even simple operations such as the vector copy operation
\begin{align*}
 \mathbf{x} \leftarrow \mathbf{y}
\end{align*}
allow for many different ways of assigning the computational work to the individual work items.
These work items are logical entities and the actual mapping to the underlying hardware is left to the OpenCL implementation.
One implementation of the kernel body for this operation with vectors of size \lstinline|N| is
\begin{lstlisting}
for (size_t i = get_global_id(0);
            i < N;
            i+= get_global_size(0))
  x[i] = y[i];
\end{lstlisting}
with \lstinline|double|-arrays \lstinline|x| and \lstinline|y|, where the vector entries are processed in blocks of the global work size given by \lstinline|get_global_size(0)|.
A drawback of the kernel above is that the increment of the index \lstinline|i| by the global work size is detrimental to caching.
As an alternative, one can rewrite the kernel in a more cache-friendly way such that each workgroup operates on consecutive memory:
\begin{lstlisting}
for (size_t i = group_start + get_local_id(0);
            i < group_end;
            i+= get_local_size(0))
  x[i] = y[i];
\end{lstlisting}
Here, \lstinline|group_start| and \lstinline|group_end| represent suitably chosen group boundaries, for example multiples of the local work size in order to preserve aligned memory accesses.

In addition to the two kernel skeletons presented above, the native OpenCL vector data types such as \lstinline|double2|, \lstinline|double4|, \lstinline|double8|, and \lstinline|double16| increase the number of possible implementations even further.
%However, these vector data types do not support strided data access well, hence have to be taken with a grain of salt.
We include these vector data types in our study even though they are less suited if strided vector operations are required.
Finally, each OpenCL kernel launch requires the specification of local and global work sizes.
We restrict the choice of local work sizes to powers of two up to a value of $512$, because other workgroup sizes are either not well-suited for parallel reduction operations such as inner products, or exhaust the available local memory.
The number of workgroups considered is empirically chosen as powers of two up to $1024$ supplemented by $48$, $80$, $96$, $112$, $160$, $192$, $224$, and $384$ in order to cover additional multiples of 16 as recommended in vendor optimization guides.
Although one may obtain negligible performance improvements with an even higher number of workgroups, we refrained from including this scenario in our benchmarks, as this would have resulted in too many configurations using large vector data types and large workgroup sizes with no work assigned.

In summary, with the parameters outlined above one obtains $2\times5\times10\times19 = 1900$ different kernel configurations.
The same parameterization and thus the same number of kernel configurations are obtained for the scaled vector addition
$\mathbf{x} \leftarrow \alpha \mathbf{y} + \beta \mathbf{z}$
with $\alpha \ne 0$ and $\beta \ne 0$, the inner product $\alpha \leftarrow \langle \mathbf{x}, \mathbf{y} \rangle$, and the dense matrix-vector product $\mathbf{x} \leftarrow \mathbf{A}\mathbf{y}$.
Although the inner product and the matrix-vector multiplication use reductions in local memory and are therefore of slightly different nature than the vector copy and addition operations, the same set of parameters can be used.
Matrix-vector products are implemented such that each workgroup computes one entry in the result vector: Workgroup \texttt{i} thus computes the inner product of row $i$ with the right hand side vector.
Other blocking strategies are possible for matrix-vector products, but as will be shown later, high performance can already be obtained in this reduced setting.

A similar parameterization is also applicable to matrix-matrix multiplications, where several additional parameters enter the optimization of (sub-)block sizes \cite{Du:CUDA-OpenCL,Goto2008,Tomov2012,Matsumoto2012,nath:matrix-matrix,Tillet:Performance-portable-linear-algebra}.
Since a detailed understanding of these additional parameters is not necessary for an understanding of the results in Section~\ref{sec:results}, we refer the interested reader to the literature at this point.

\section{Results} \label{sec:results}

The 1900 kernel configurations for each of the four operations above was compared on different hardware from AMD, INTEL, and NVIDIA, using the OpenCL implementation of the respective vendor.
We selected both older and newer graphics processing units (GPUs) from AMD and NVIDIA in order to quantify the influence of more recent hardware on performance portability.
Both the AMD Radeon HD 5850 (with Catalyst 14.1 on Windows 7) and the NVIDIA GeForce GTX 285 (with CUDA 5.0) were introduced in 2009, while the AMD FirePro W9000 (with Catalyst Pro 13.251.1) and the NVIDIA Tesla K20m (with CUDA 6.0) were released as high-end workstation GPUs in 2012 and 2013, respectively.
Furthermore, results for the central processing unit (CPU) of an AMD A10-5800K accelerated processing unit (with AMD APP SDK 2.8 on Windows 7), a dual-socket INTEL Xeon E5-2670 machine, and an INTEL Xeon Phi are presented (both with INTEL OpenCL SDK 2013 XE 3.0).
The local workgroup size was limited to 256 work items on AMD hardware, because larger sizes are not supported.
As we were unable to obtain satisfactory performance for memory bandwidth limited operations using OpenCL on the INTEL Xeon Phi, we mostly plot results for the CPUs and GPUs in our comparison.
This performance regression has been confirmed by INTEL engineers in direct communication for the case of vector addition and has also been reported by other colleagues in private communication.
%Although we collected test results on other hardware from these vendors as well, this selection resembles all important features observed during our studies.

Double precision arithmetic was used for all benchmarks, since this is the standard for linear algebra operations.
Because all operations considered are limited by memory bandwidth, all results are presented relative to the theoretical peak memory bandwidth of the respective device to simplify comparisons.
For matrix-vector products we assume perfect caching of the right hand side vector and only counted the minimum number of bytes necessary to compute the results.
With the large caches on CPUs and broadcast memory transfers on GPUs, this simplification is justified and results in consistent performance results on most hardware considered.
To ensure that the PCI-Express latency of up to 10 microseconds is negligible, that full data caching is not possible, and that the benchmark can also be run on GPUs with only 512 MB of main memory, vectors for the copy, addition, and inner product operations are chosen to consist of two million entries each.
Similarly, the row and column counts of the matrix for the matrix-vector product are 2048.

%%%%%%%%%%%%%%%%%%%%%%%%%%%%%%%%%% Feature Extraction %%%%%%%%%%%%%%%%%%%%%%%%%%%%%%%%%%%%%%%

\subsection{Isolated Parameter Variation}
We first consider benchmark results obtained by looking at the distribution of performances for each of the four parameters identified in the previous section:
\begin{itemize}
 \item \textbf{Increment Type}: Whether the loop index is incremented by the local work size (\emph{local}) or by the global work size (\emph{global}).
 \item \textbf{Vector Length}: Whether to use the scalar data type \lstinline|double| (vector length 1), or native vector types such as \lstinline|double16|.
 \item \textbf{Local Work Size}: The number of work items per OpenCL workgroup.
 \item \textbf{Workgroups}: Total number of OpenCL workgroups.
\end{itemize}
The benchmark results for the four operations considered are qualitatively similar, with differences being more pronounced for the inner product and the matrix-vector product kernels.
Consequently, we only depict results for the dot product in the following.

\begin{figure*}
 \centering
 \includegraphics[width=0.31\textwidth]{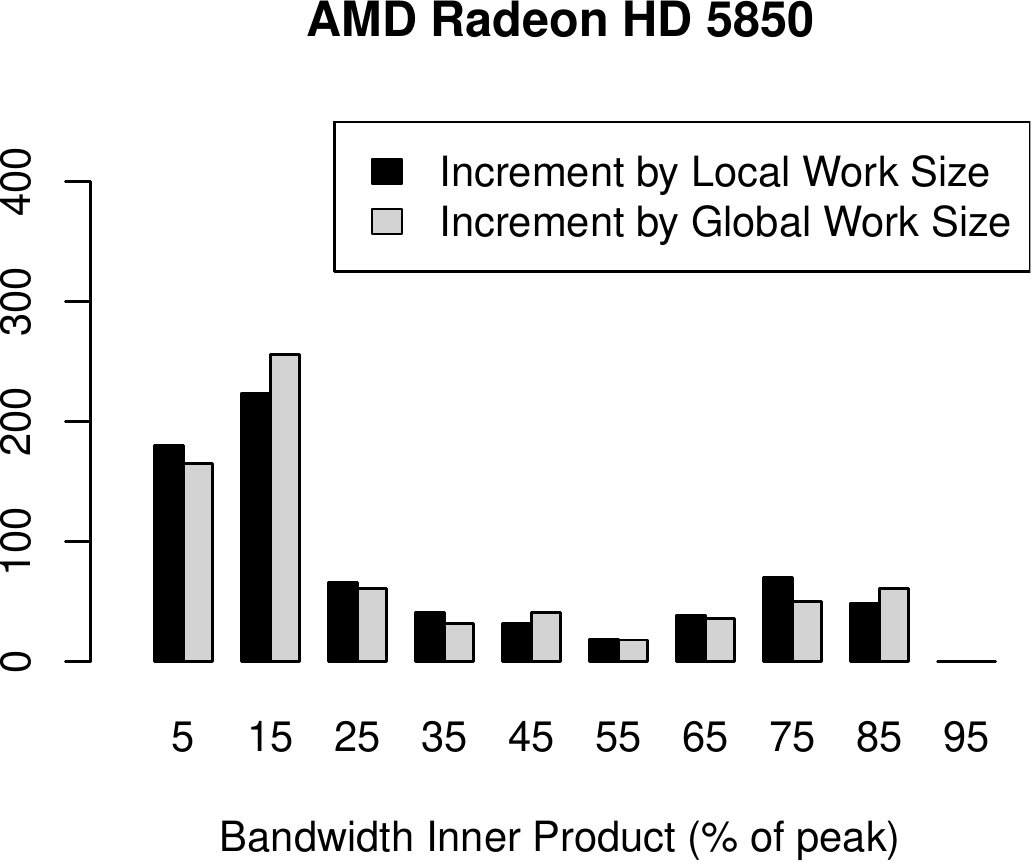} \hfill
 \includegraphics[width=0.31\textwidth]{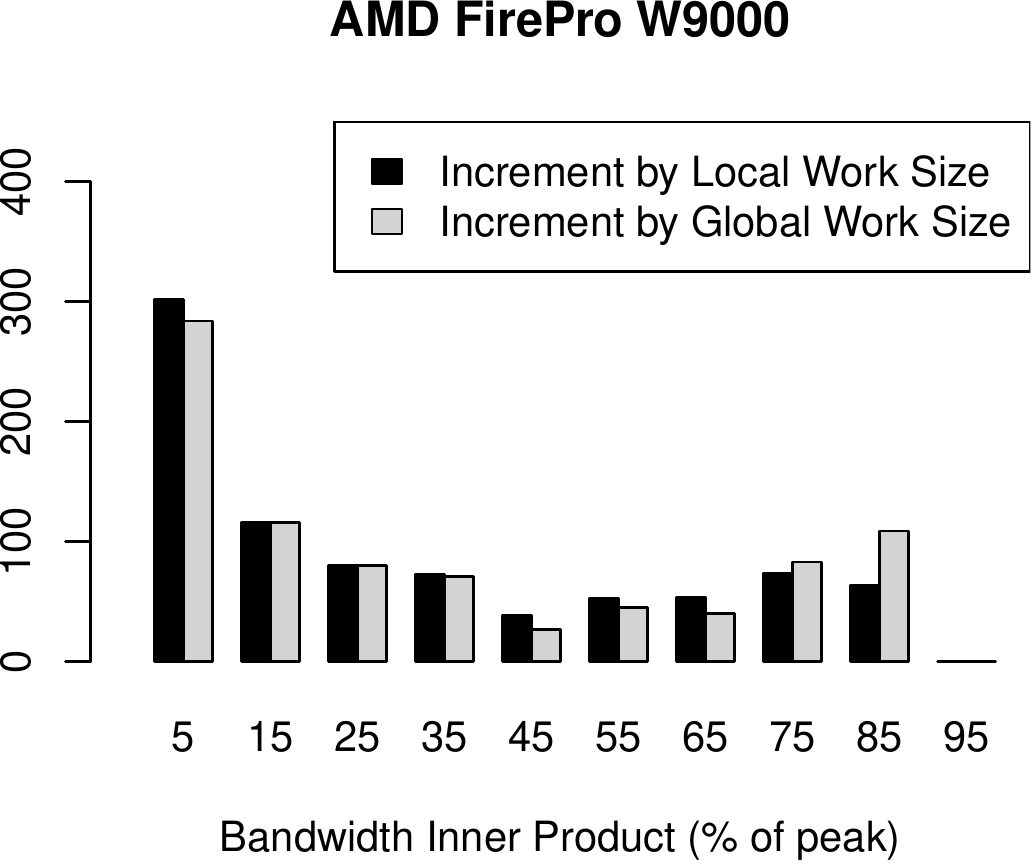} \hfill
 \includegraphics[width=0.31\textwidth]{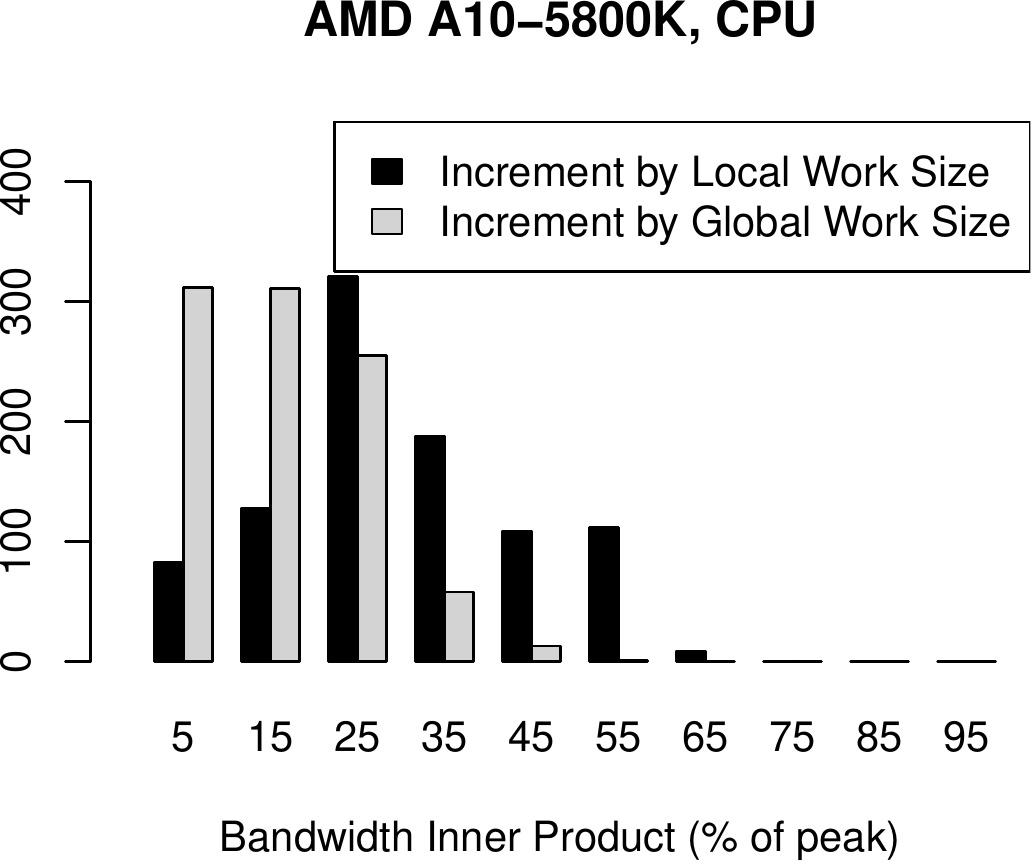} \\[0.5em]
 \includegraphics[width=0.31\textwidth]{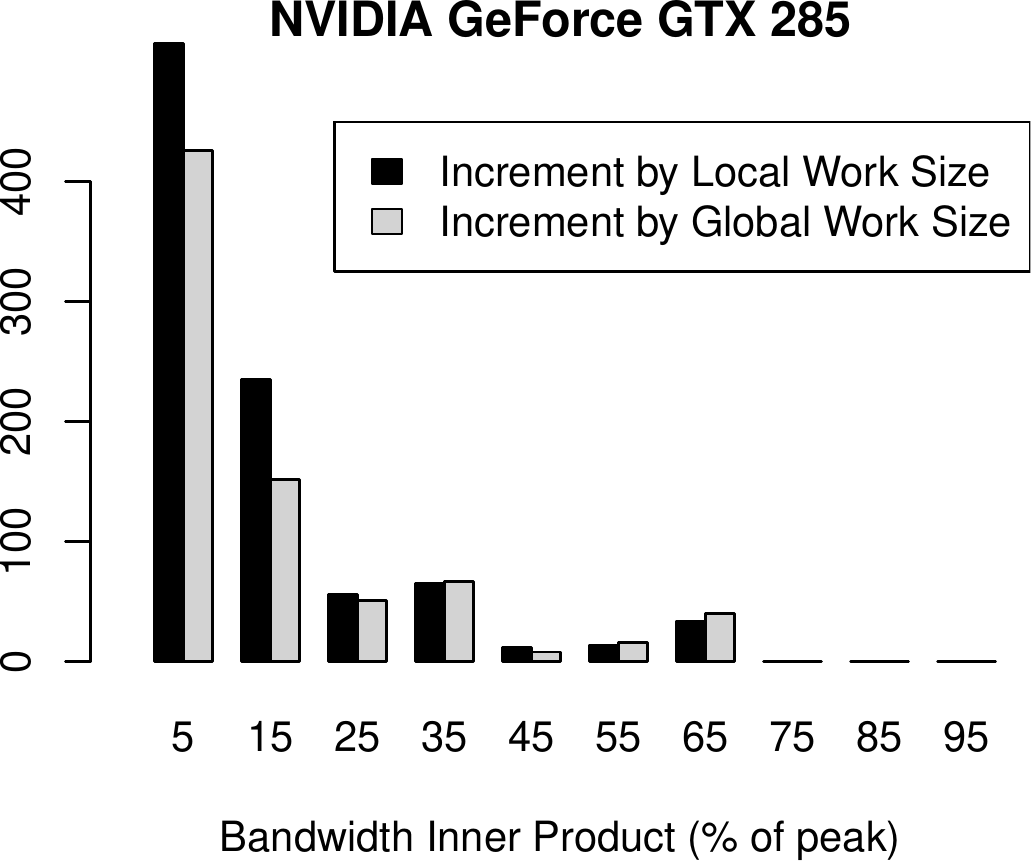} \hfill
 \includegraphics[width=0.31\textwidth]{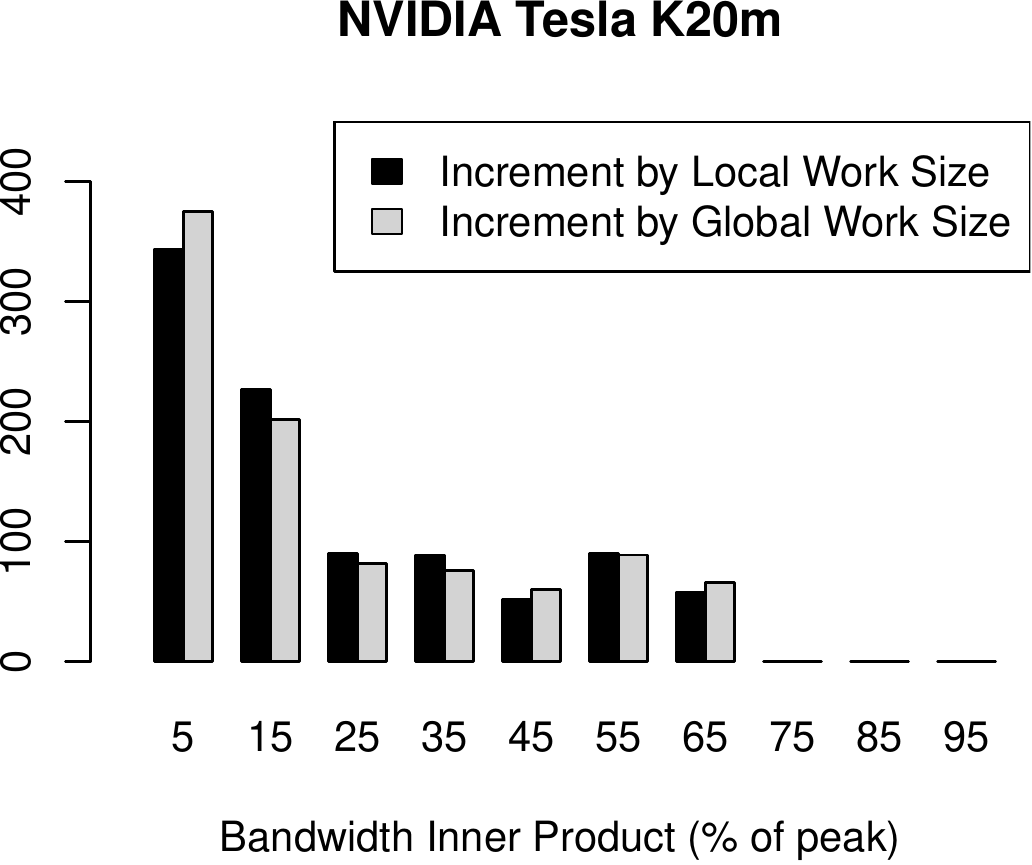} \hfill
 \includegraphics[width=0.31\textwidth]{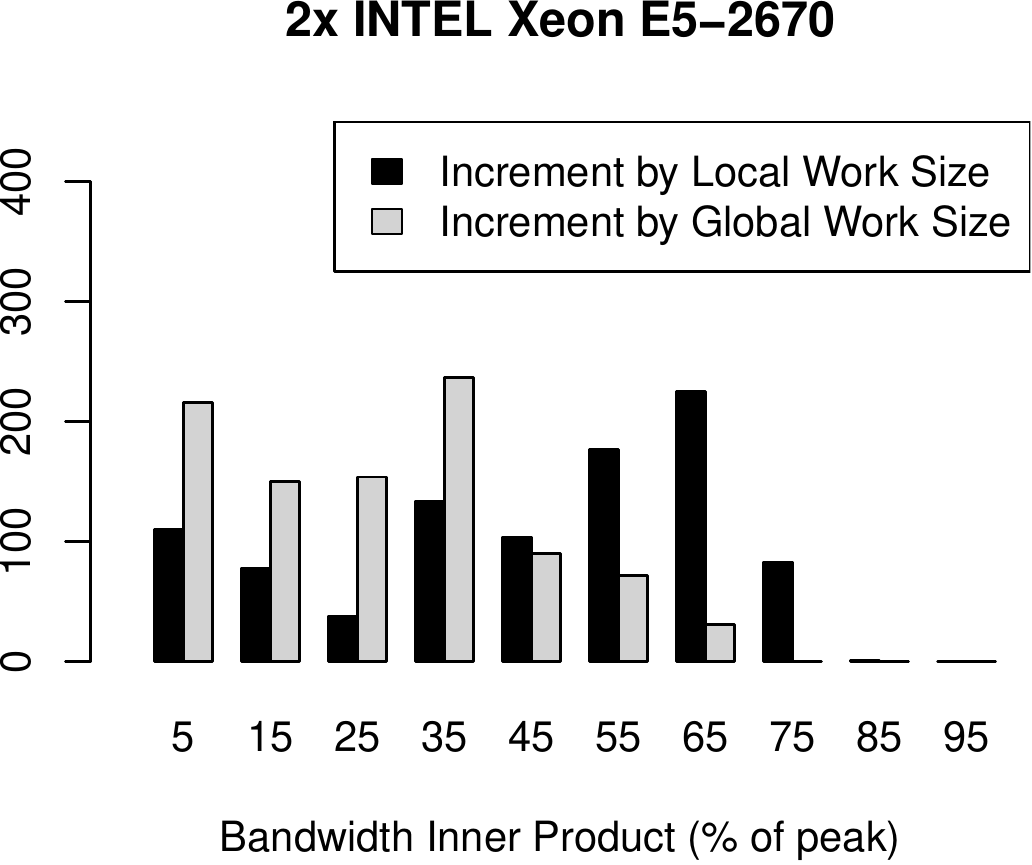} \\
 \caption{Frequency plots of the 1900 kernel configurations for the inner product operation.
          An increment by the local workgroup size is favorable over an increment by the global workgroup size for CPUs, here shown for a dual-socket INTEL Xeon E5-2670 machine.
          On GPUs an increment by the global workgroup size results in slightly better performance.}
 \label{fig:increment-plots}
\end{figure*}

Figure~\ref{fig:increment-plots} shows frequency plots with respect to the increment type in the \lstinline|for|-loop.
While the increment type has only a very mild influence on GPUs, there is a clear preference of increments by the local work size on CPUs, which is attributed to memory pages loaded from main memory.
Therefore, if no information over the target device is available other than the device type, which can be obtained from the OpenCL stack at runtime, an increment by the global work size should be used for GPUs, and an increment by the local work size for CPUs.

\begin{figure*}
 \centering
 \includegraphics[width=0.31\textwidth]{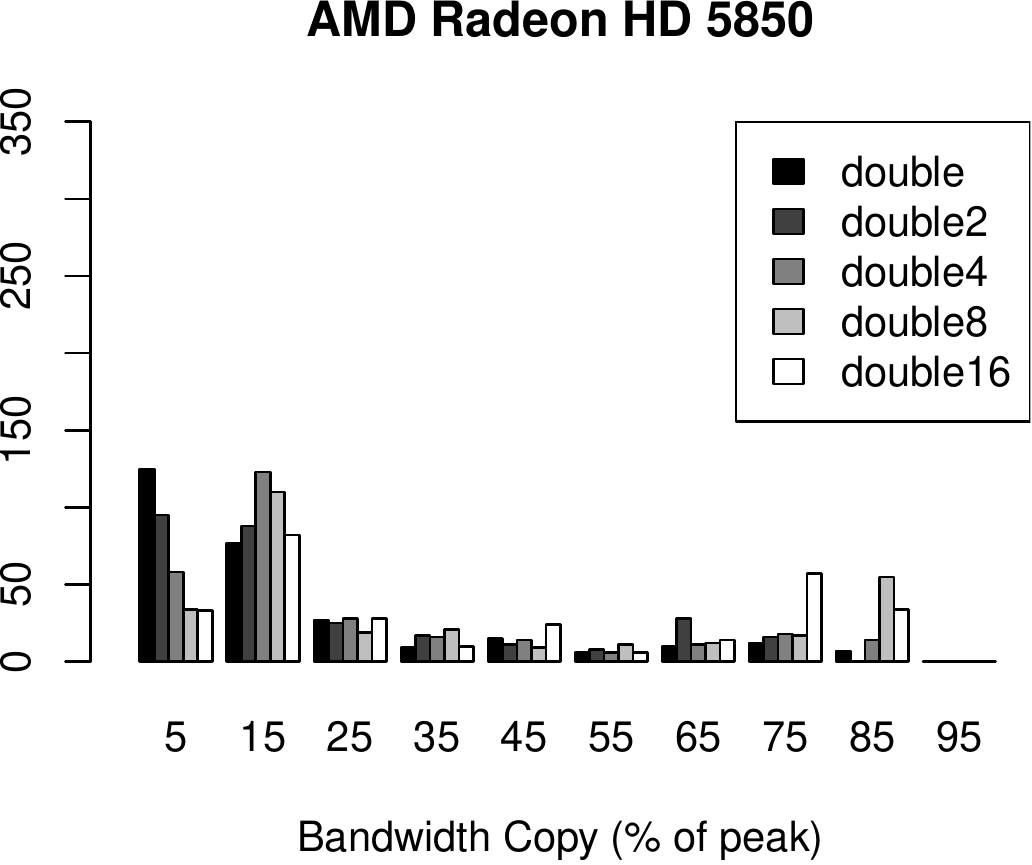} \hfill
 \includegraphics[width=0.31\textwidth]{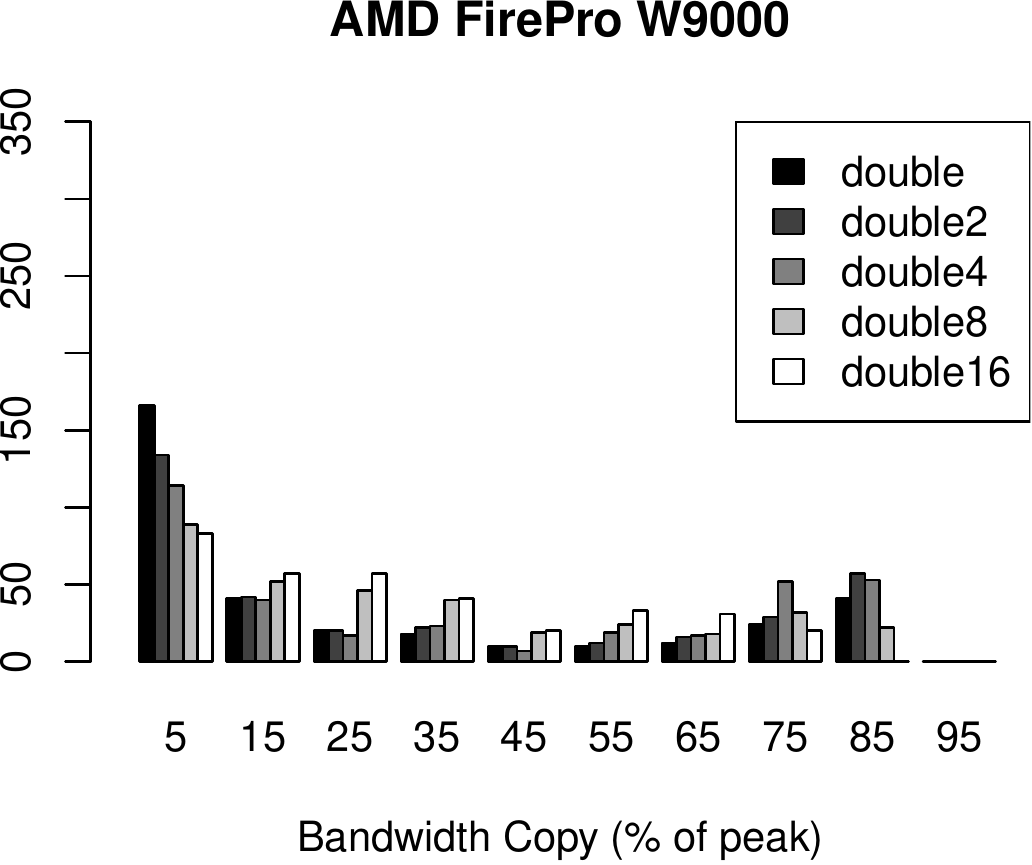} \hfill
 \includegraphics[width=0.31\textwidth]{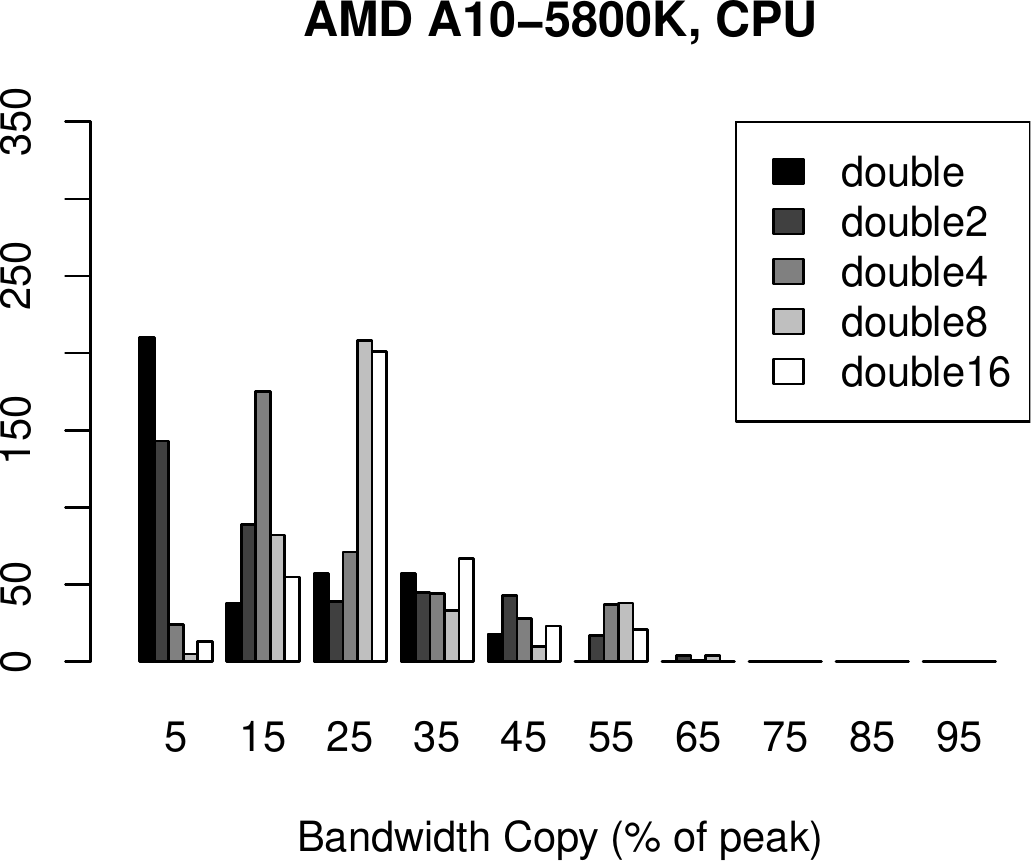} \\[0.5em]
 \includegraphics[width=0.31\textwidth]{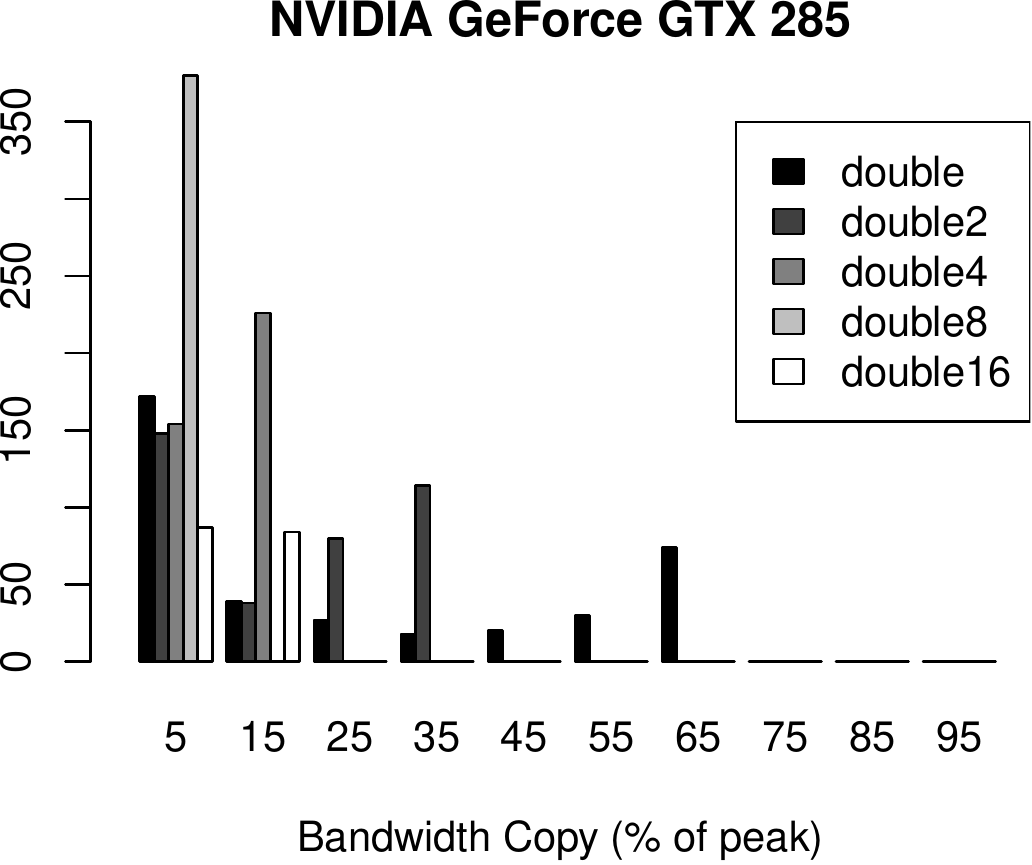} \hfill
 \includegraphics[width=0.31\textwidth]{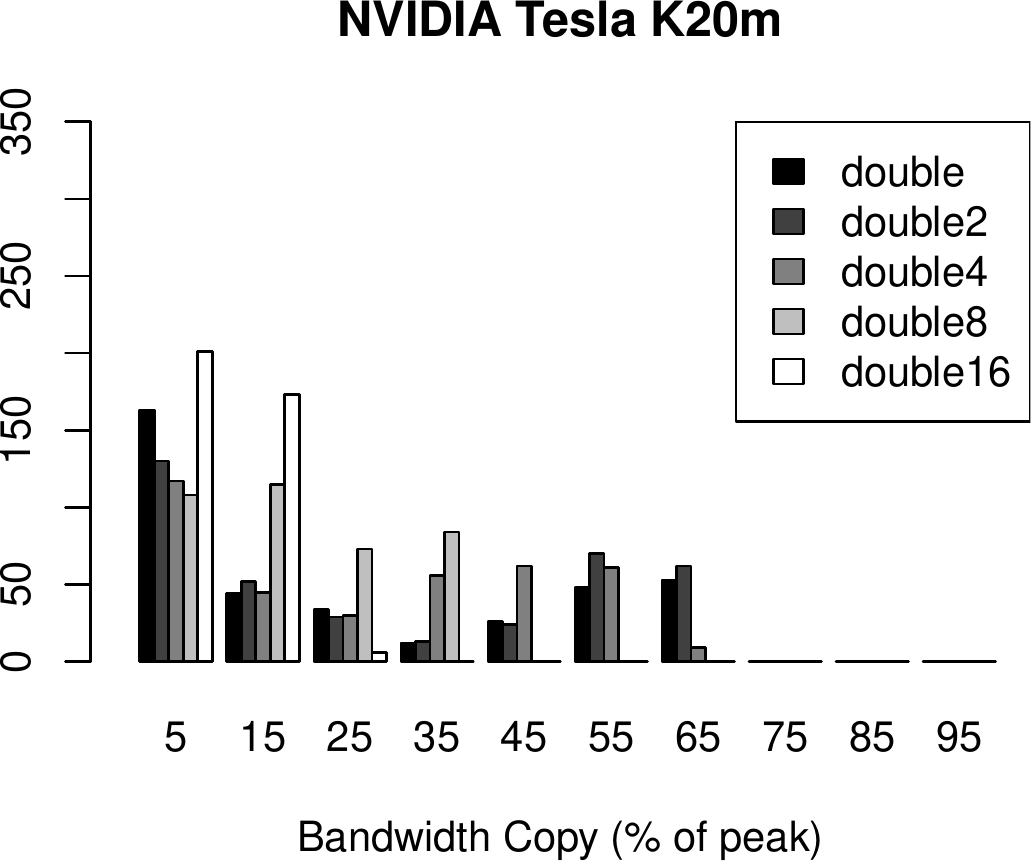} \hfill
 \includegraphics[width=0.31\textwidth]{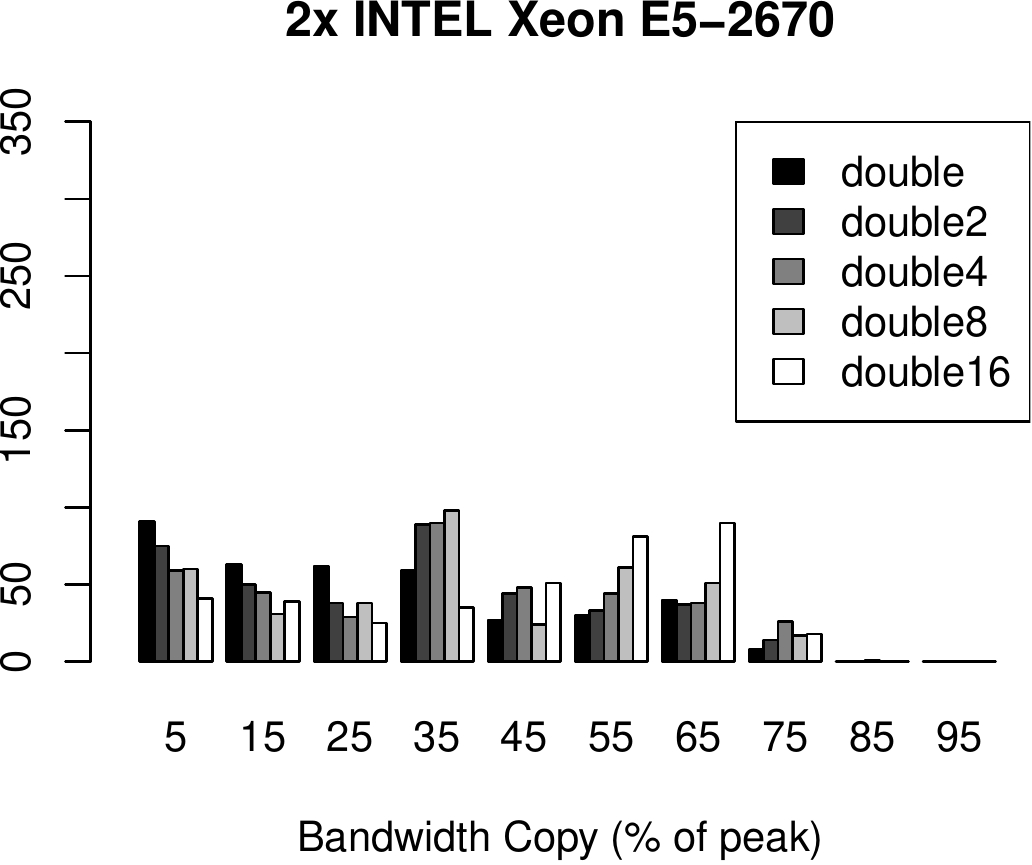} \\
 \caption{Frequency plots of the 1900 kernel configurations for the inner product operation investigating different vector data types.
          Although vector data types other than \lstinline|double| occasionally result in higher performance, \lstinline|double| shows the highest overall performance portability.
          }
 \label{fig:vec-plots}
\end{figure*}

A comparison of benchmark results for the different vector data types is given in Figure~\ref{fig:vec-plots}.
The HD 5850 prefers longer vector types, particularly \lstinline|double8|.
Conversely, vector data types result in poor performance on the GTX 285, which prefers the plain \lstinline|double| scalar type for its scalar architecture.
On the W9000 and K20m one can obtain good performance with both \lstinline|double|, \lstinline|double2| and to a lesser extent \lstinline|double4|, while \lstinline|double8| and \lstinline|double16| result in poor performance on the K20m.
CPUs are less sensitive to the different vector types, with \lstinline|double4| being slightly preferred on the Xeon E5-2670, whereas \lstinline|double| is not recommended on the A10-5800K CPU.

\begin{figure*}
 \centering
 \includegraphics[width=0.31\textwidth]{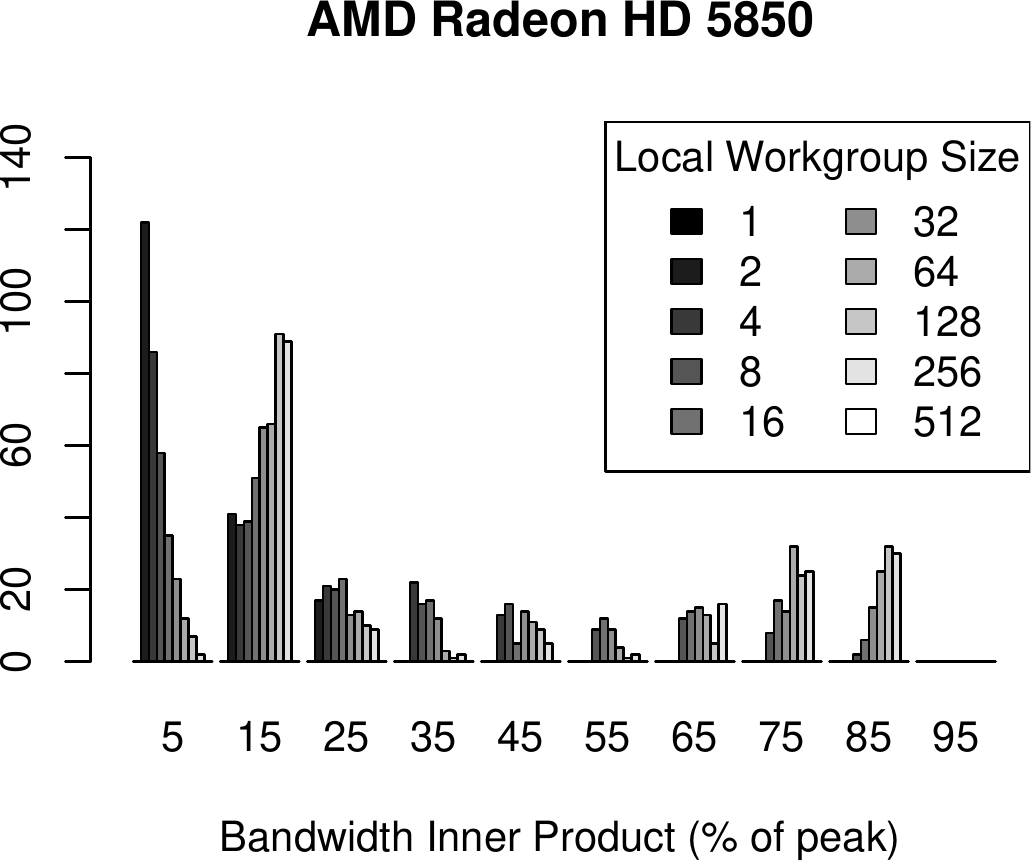} \hfill
 \includegraphics[width=0.31\textwidth]{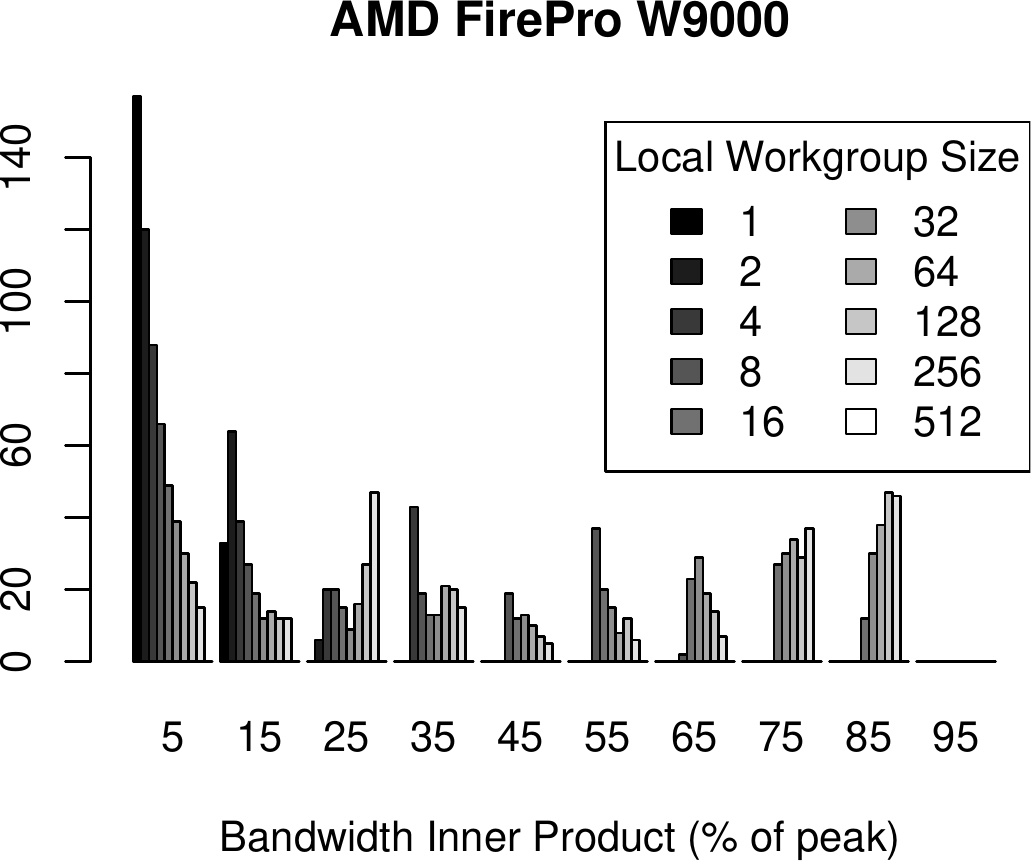} \hfill
 \includegraphics[width=0.31\textwidth]{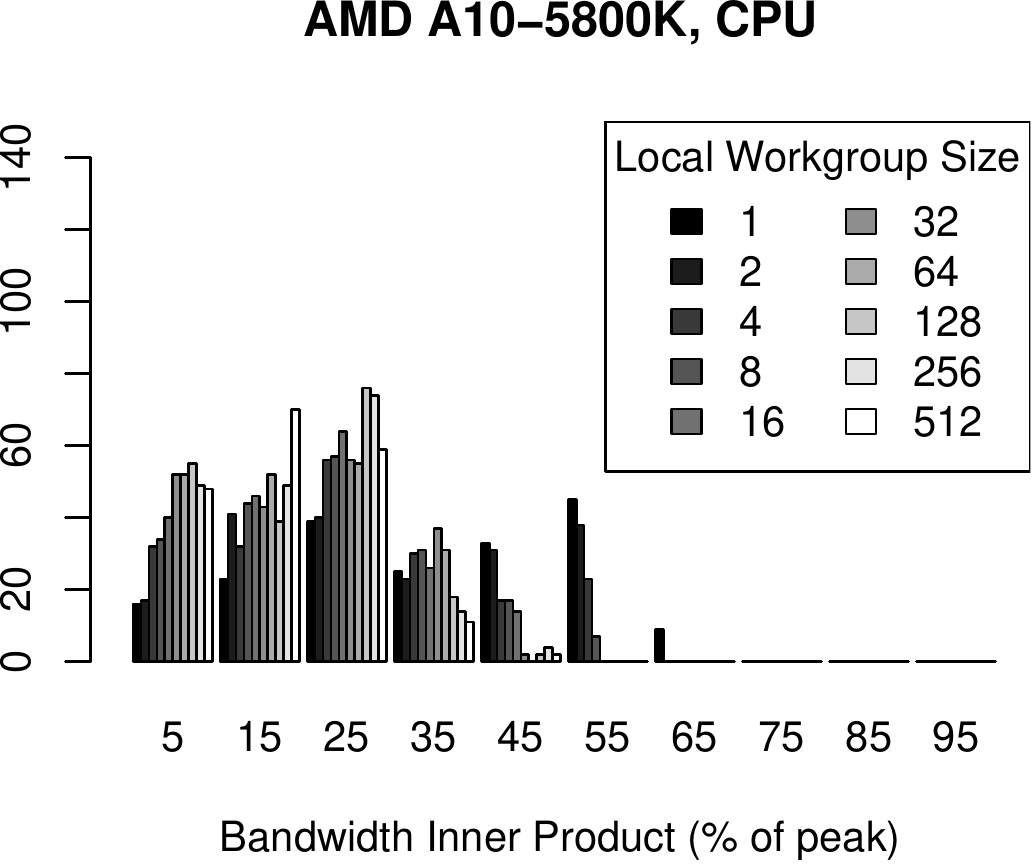} \\[0.5em]
 \includegraphics[width=0.31\textwidth]{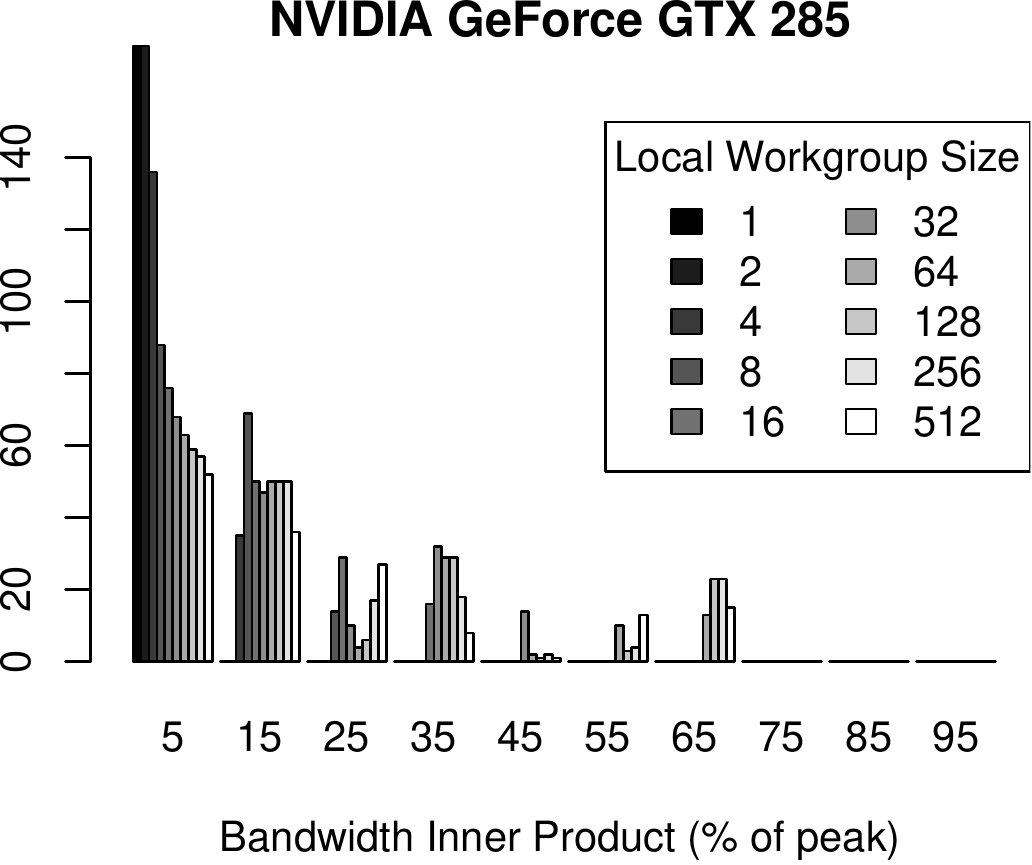} \hfill
 \includegraphics[width=0.31\textwidth]{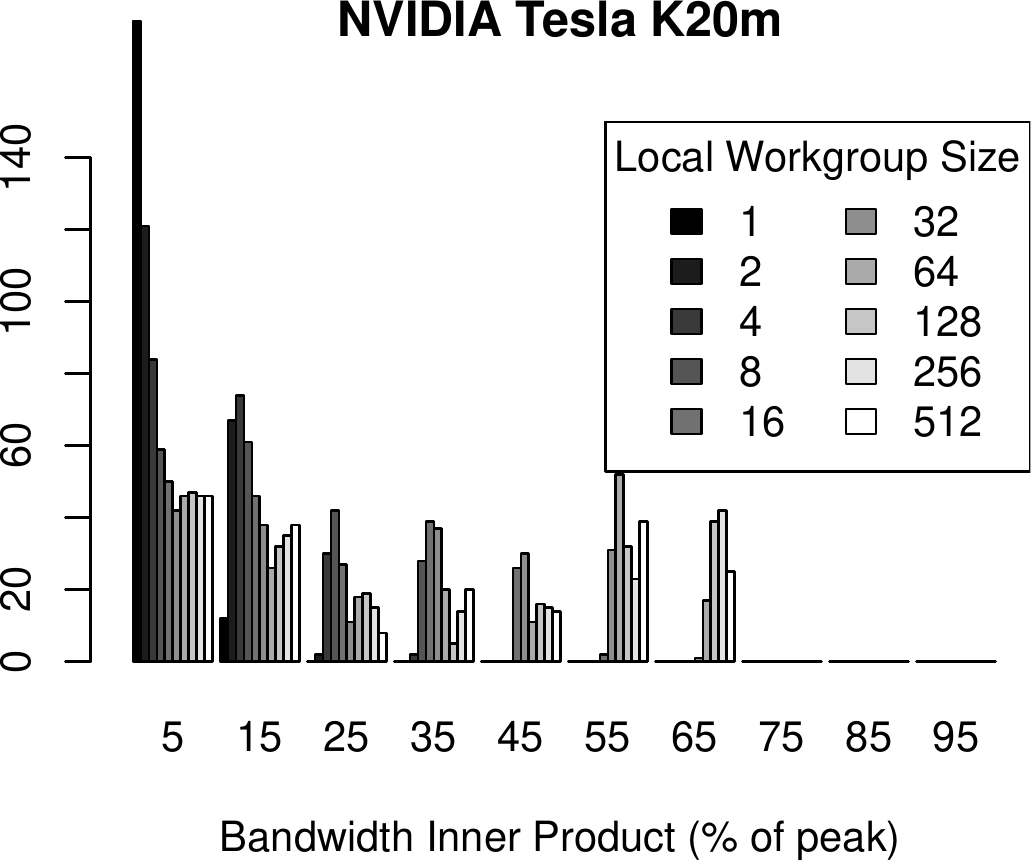} \hfill
 \includegraphics[width=0.31\textwidth]{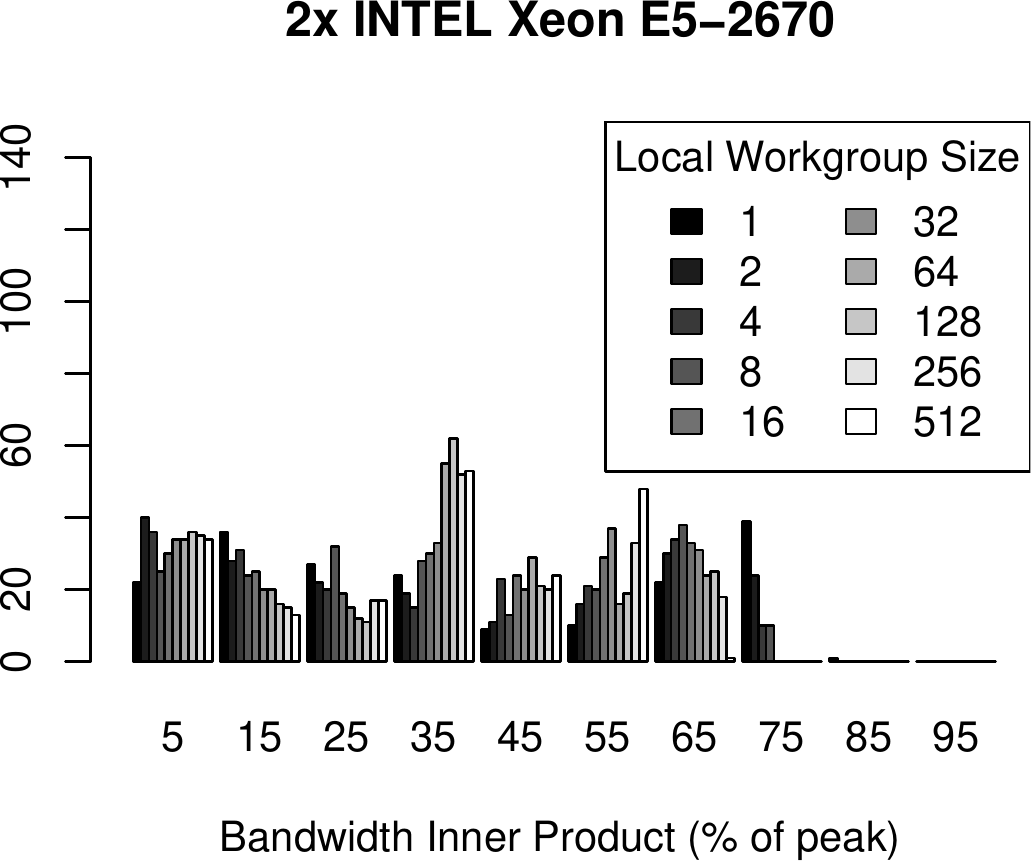} \\
 \caption{Frequency plots of the 1900 kernel configurations for the inner product operation investigating different local workgroup sizes.
          Local workgroup sizes of 128 and 256 provide the best overall performance on GPUs, while a local workgroup size of 1 is best for CPUs.}
 \label{fig:localws-plots}
\end{figure*}

Figure~\ref{fig:localws-plots} depicts the benchmark results obtained for different local workgroup sizes.
Both AMD and NVIDIA GPUs show good performance for 128 and 256 work items per workgroup.
In contrast, good performance on the CPU is predominantly obtained with only one or two work items per workgroup.
The same trend is observed for matrix-vector products, while large workgroup sizes yield better results for copy and addition (plots not shown).
The INTEL CPUs are found to work slightly better with large workgroup sizes than the AMD CPU, which barely reaches more than half of the theoretical peak memory bandwidth.
We suspect that this is caused by the use of only a single memory channel by the OpenCL implementation.

\begin{figure*}
 \centering
 \includegraphics[width=0.31\textwidth]{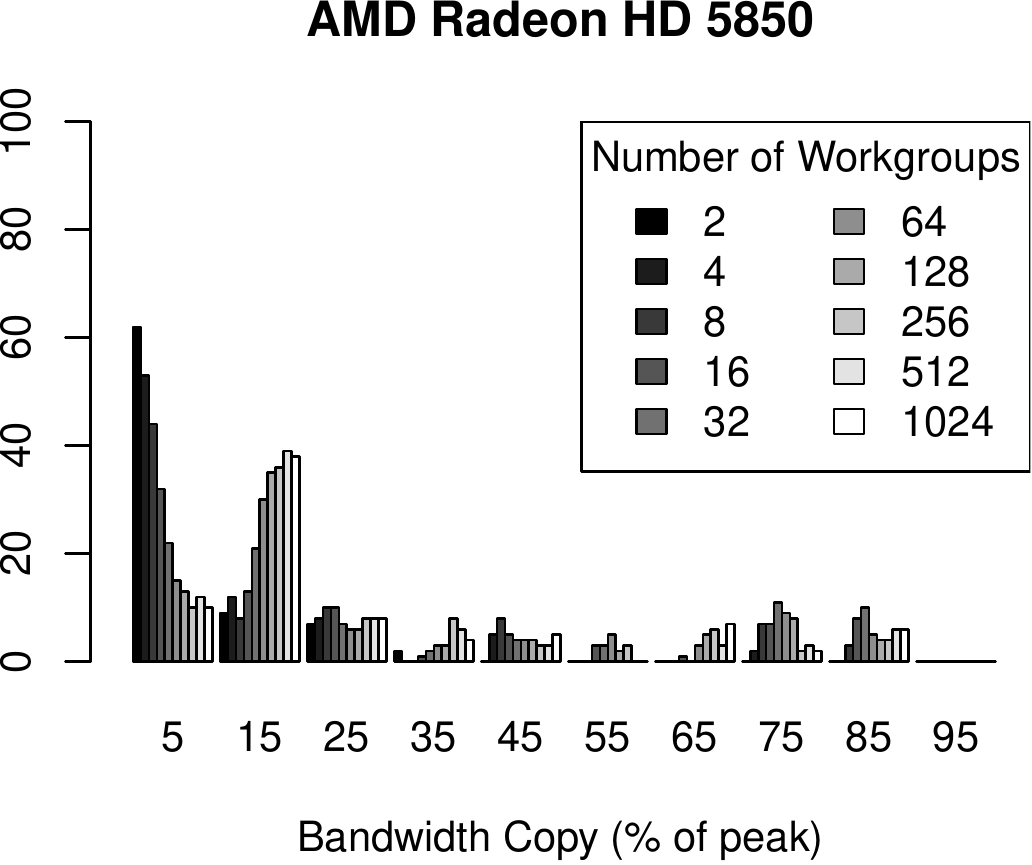} \hfill
 \includegraphics[width=0.31\textwidth]{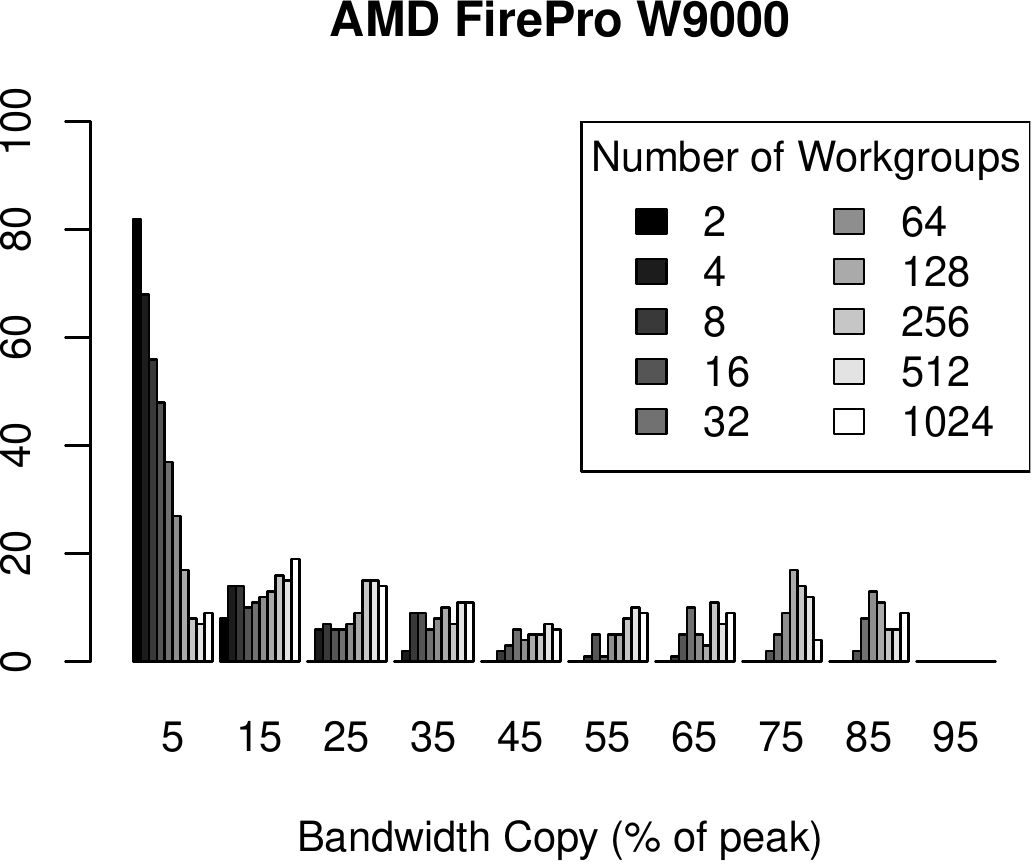} \hfill
 \includegraphics[width=0.31\textwidth]{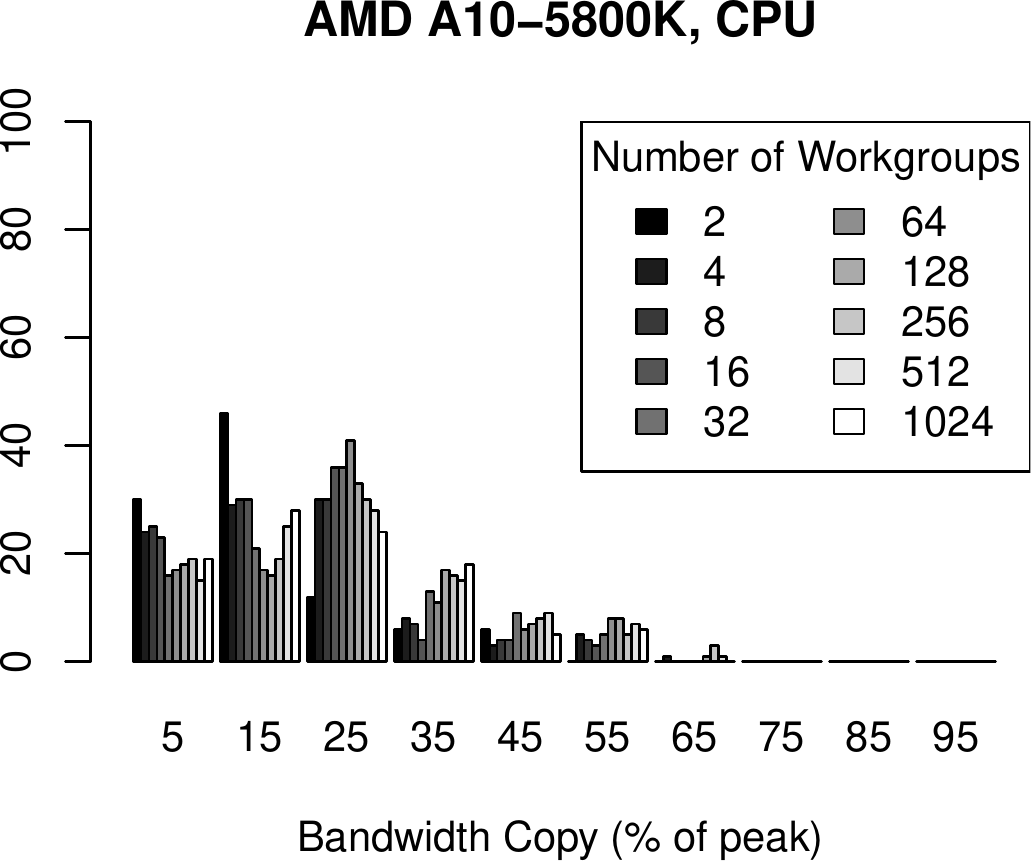} \\[0.5em]
 \includegraphics[width=0.31\textwidth]{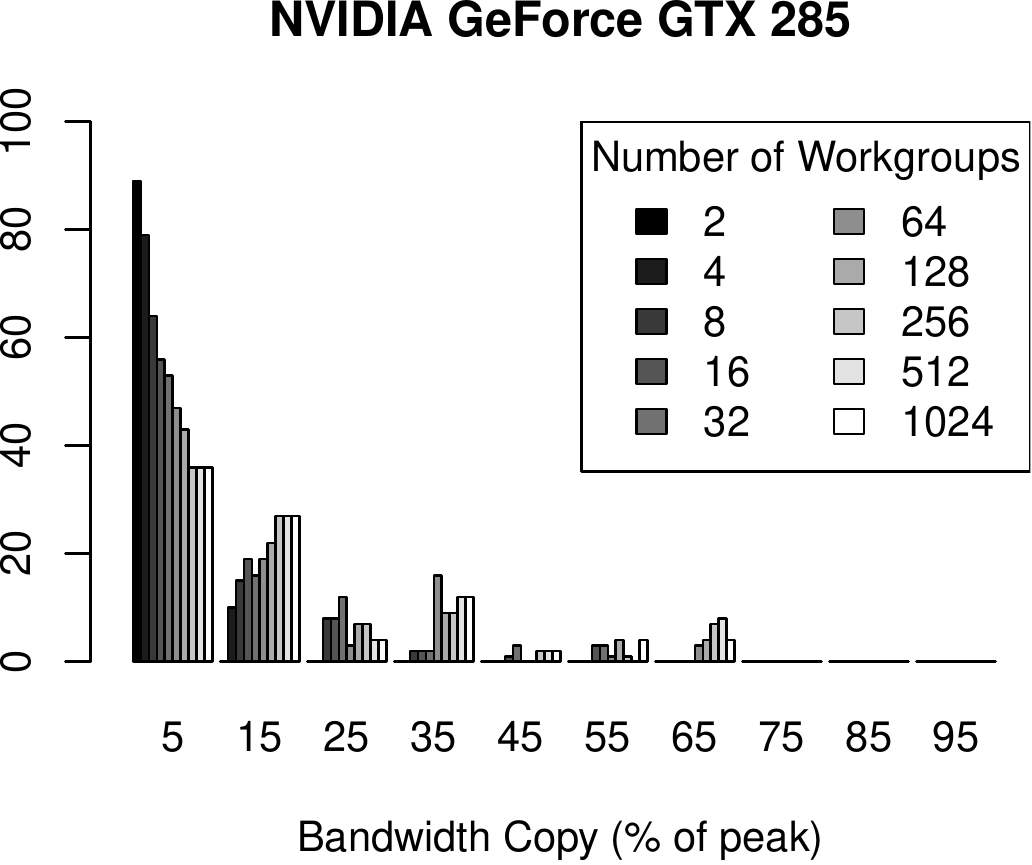} \hfill
 \includegraphics[width=0.31\textwidth]{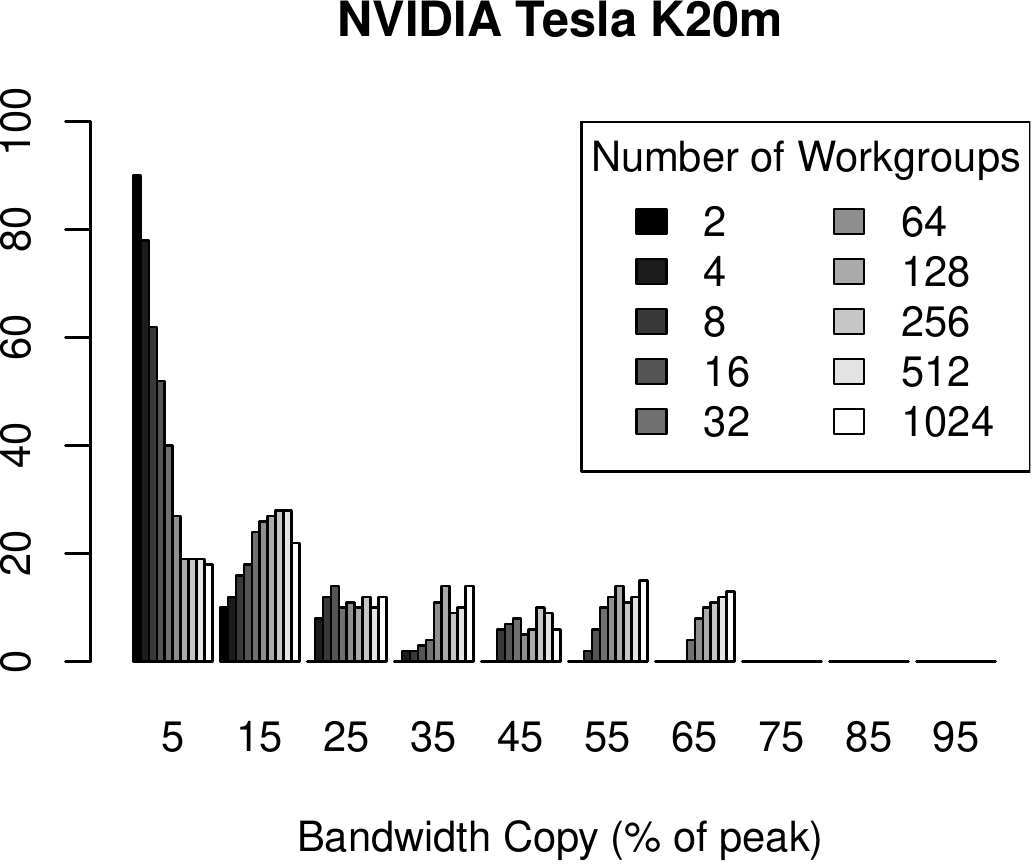} \hfill
 \includegraphics[width=0.31\textwidth]{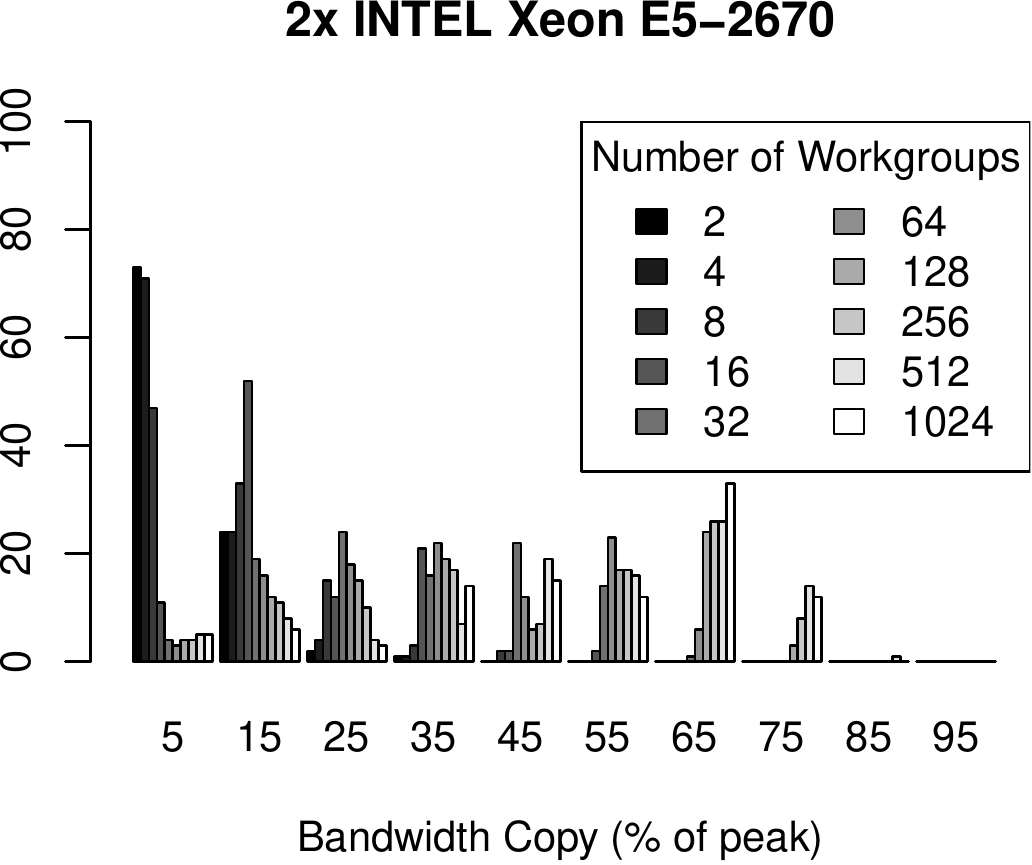} \\
 \caption{Frequency plots of the 1900 kernel configurations for the inner product operation investigating different number of workgroups.
          In contrast to other device types and GPUs from NVIDIA, GPUs from AMD do not necessarily favor a higher number of workgroups.}
 \label{fig:wgroups-plots}
\end{figure*}

Finally, Figure~\ref{fig:wgroups-plots} shows the benchmark results with respect to different numbers of workgroups, where only powers of two are plotted for the sake of conciseness.
Good performance on AMD GPUs can be obtained already with $32$ workgroups, whereas the results for NVIDIA GPUs as well as the CPUs show a clear preference for a high number of workgroups.
We conclude that the number of workgroups should generally be chosen as high as reasonably possible particularly on NVIDIA GPUs and no detrimental effects on performance are to be expected by this choice.

%%%%%%%%%%%%%%%%%%%%%%%%%%

\subsection{Portability Within Each Device}

% GPUs:
\begin{figure*}
 \centering
 \subfigure{ \includegraphics[width=0.945\textwidth]{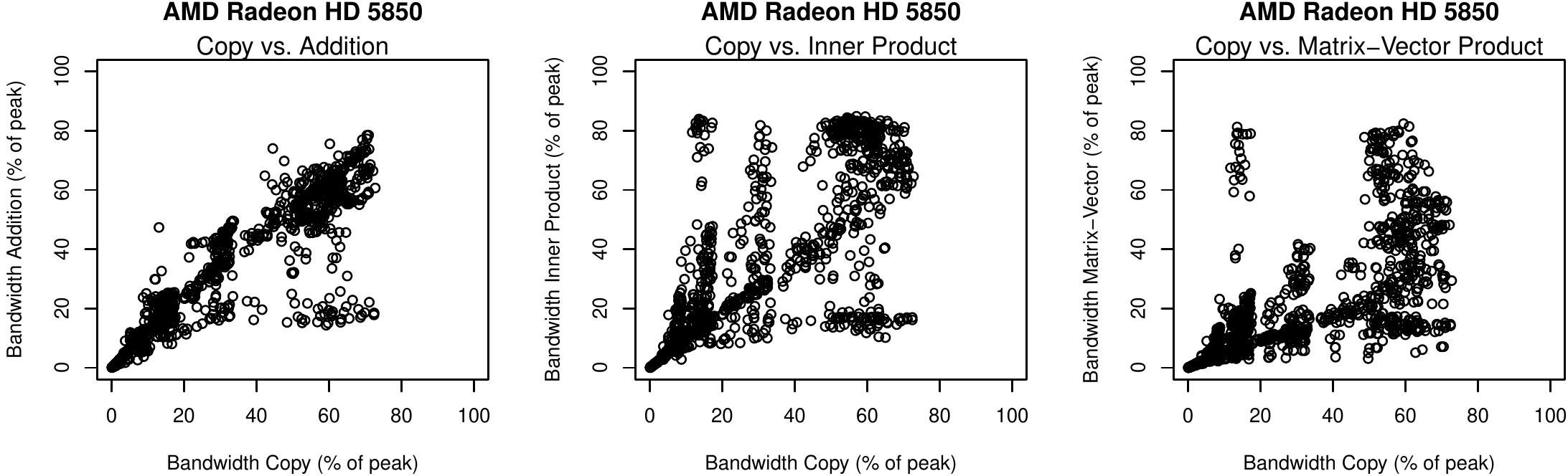} }
 \subfigure{ \includegraphics[width=0.945\textwidth]{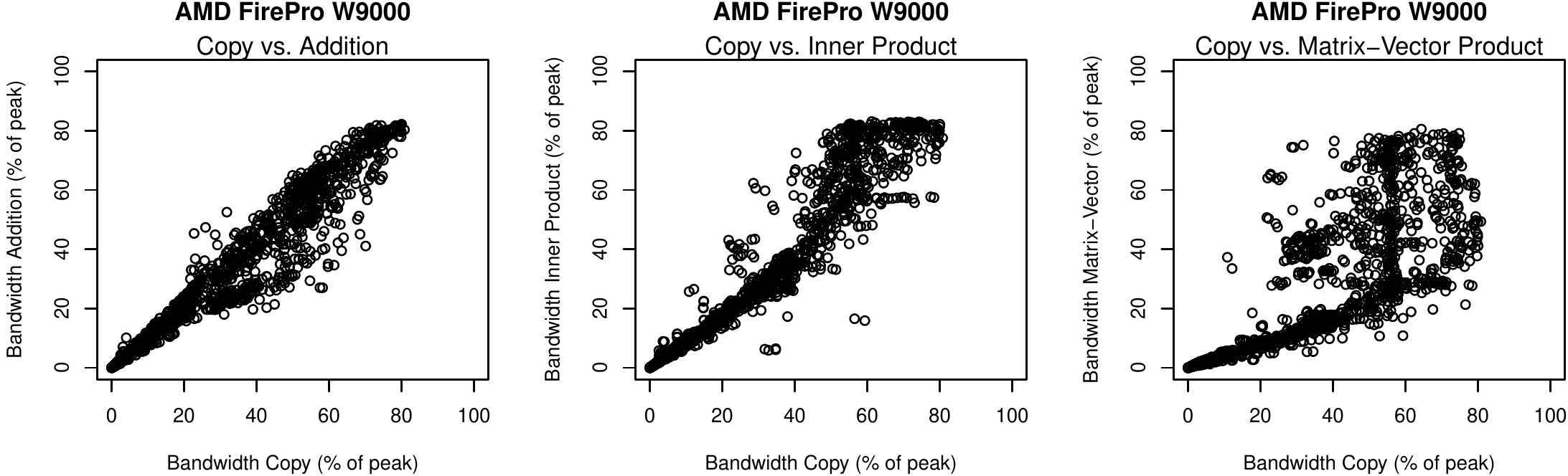} }
 \subfigure{ \includegraphics[width=0.945\textwidth]{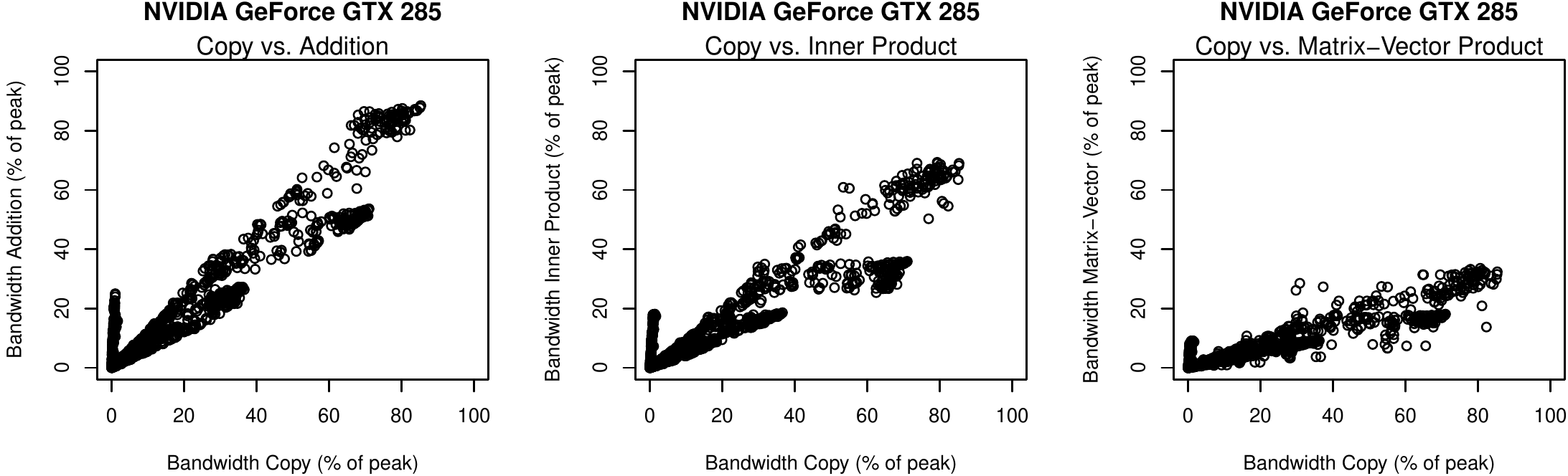} }
 \subfigure{ \includegraphics[width=0.945\textwidth]{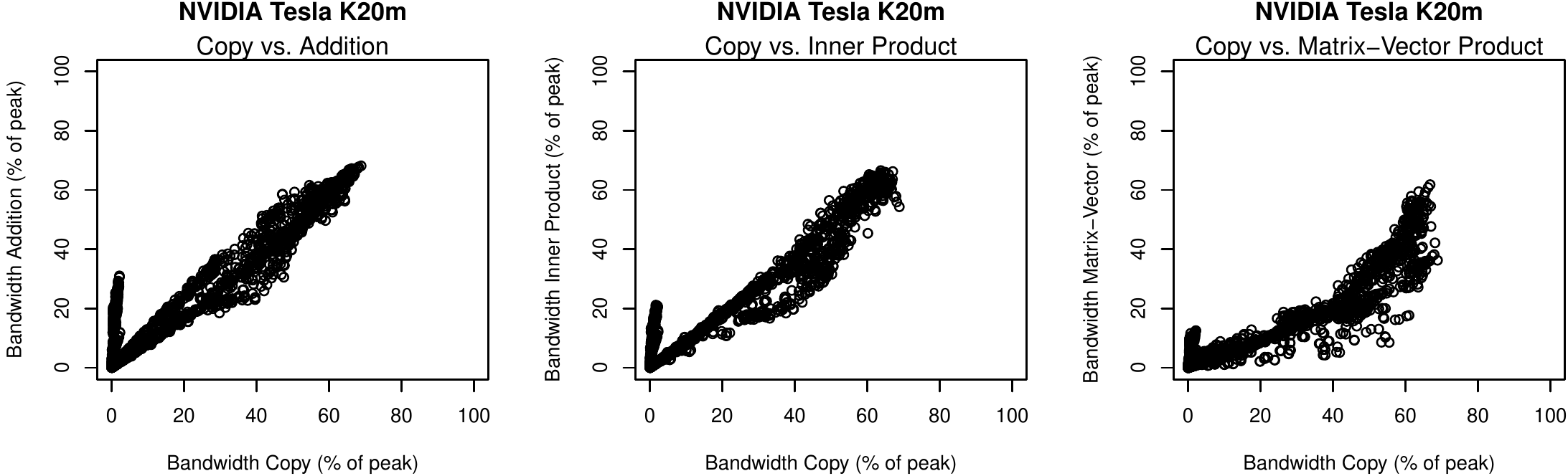} }
 \caption{Comparison of kernel performances for GPUs from AMD and NVIDIA.
          Good performance for the copy kernel usually implies high performance for the other kernels on NVIDIA GPUs.
          Kernel configurations with high performance for the copy kernel are candidates for good performance of matrix-vector products on AMD GPUs.}
 \label{fig:homogeneous-device-plots-gpu}
\end{figure*}

% CPUs
\begin{figure*}
 \centering
 \subfigure{ \includegraphics[width=0.99\textwidth]{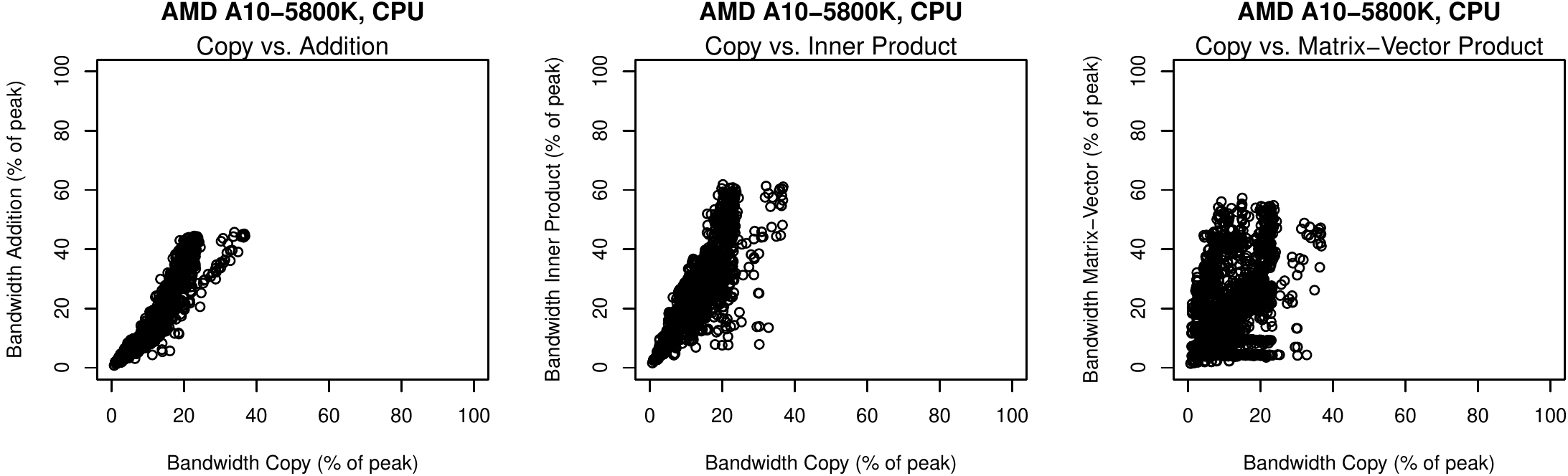} }
 \subfigure{ \includegraphics[width=0.99\textwidth]{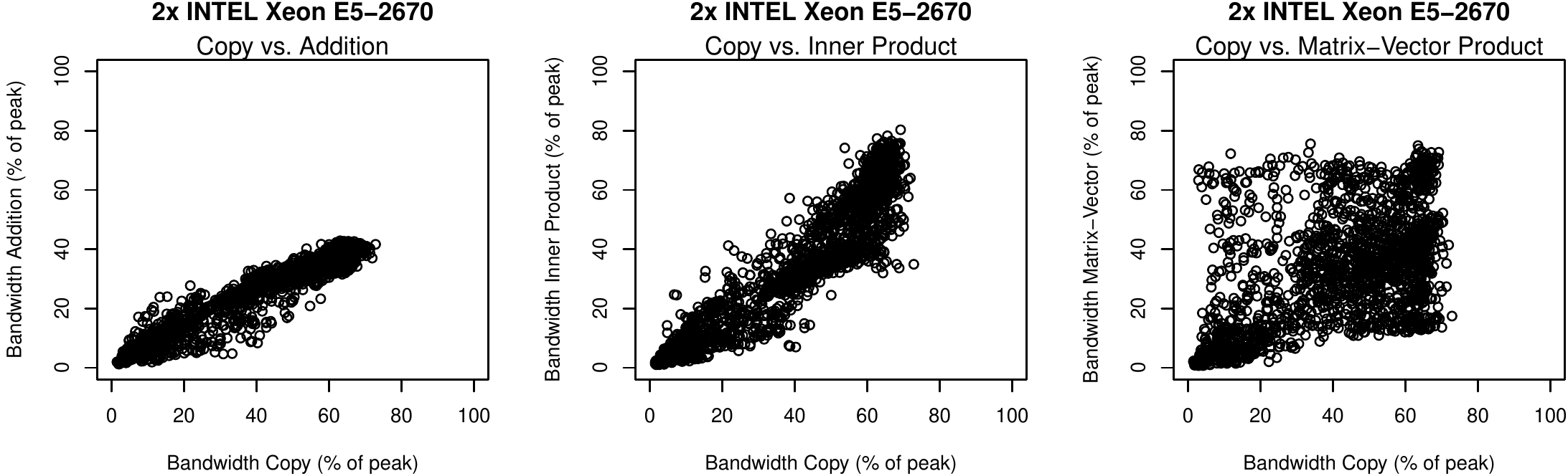} }
 \subfigure{ \includegraphics[width=0.99\textwidth]{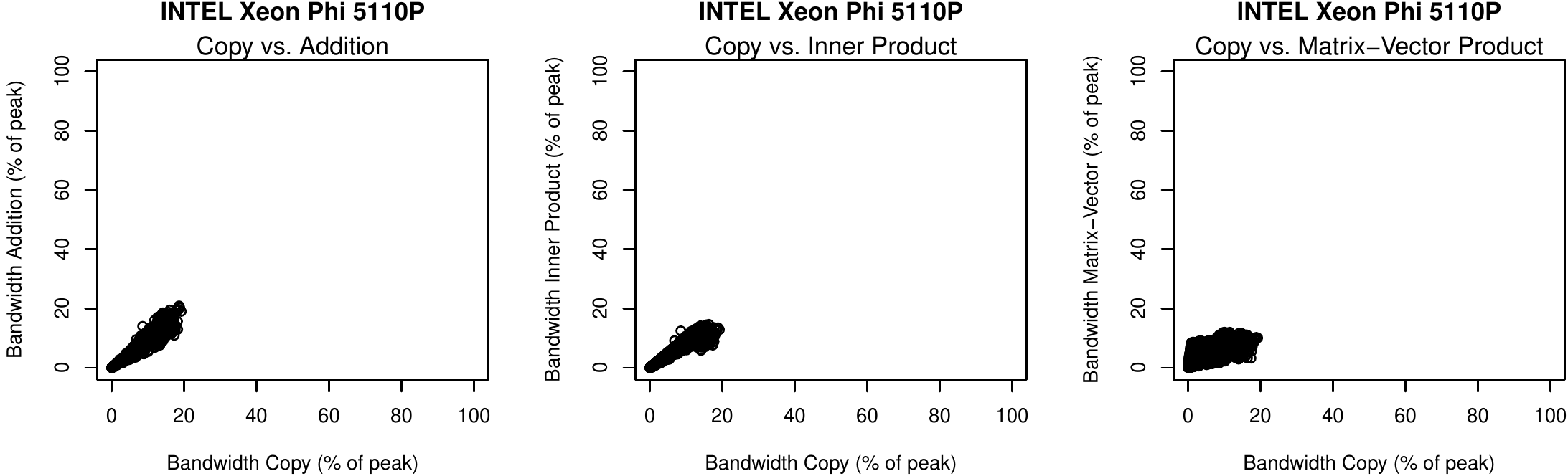} }
 \caption{Comparison of kernel performances on an AMD A10-5800K CPU, a dual-socket INTEL Xeon E5-2670 machine, and the INTEL Xeon Phi.
          On AMD GPUs and the Intel CPUs one can use the copy kernel to obtain candidate kernel configurations for other operations.
          The performance correlation is smaller than for GPUs.
          Performance on the INTEL Xeon Phi when using OpenCL remains poor even after extensive autotuning.}
 \label{fig:homogeneous-device-plots-cpu}
\end{figure*}

Figure~\ref{fig:homogeneous-device-plots-gpu} and Figure~\ref{fig:homogeneous-device-plots-cpu} illustrate the change in performance of each kernel configuration for the four operations, always taking the performance obtained for the copy kernel as reference on the abscissa.
If the performance obtained for the copy operation were the same for the other three operations, then all configurations would be aligned along the diagonal.
The best configurations are located in the upper right corners, indicating that they result in high performance for both operations compared.

NVIDIA GPUs show a strong correlation of performance, where a good performance for the copy operation also results in good performance for the addition, the inner product, and the matrix-vector product, cf.~Figure~\ref{fig:homogeneous-device-plots-gpu}.
The overall performance drop for matrix-vector products on the GTX 285 can be attributed to the absence of caching for the right hand side vector.
Good performance correlation is also obtained on the W9000, even though there is a higher performance fluctuation when comparing the copy operation with the matrix-vector product.
Nevertheless, by considering only configurations achieving more than $75$ percent of the theoretical peak bandwidth for the copy kernel, configurations offering more then $75$ percent of theoretical peak bandwidth for the matrix-vector product are included.
The weakest performance correlation among the GPUs was found for the HD 5850: For example, configurations exist which obtain $75$ percent of peak memory bandwidth for the copy operation, but only $20$ percent of peak memory bandwidth for the addition despite of the strong similarity of the two operations.
Regardless, by considering only configurations with more than $60$ percent of peak memory bandwidth for the copy operation, high performance can also be obtained for the other operations.

Figure ~\ref{fig:homogeneous-device-plots-cpu} show that a similar pattern holds true for CPUs, yet the correlation of performances for the different operations is particularly weak for the matrix-vector product.
Still, the search space for tuning these operations can be substantially reduced to only those configurations providing good bandwidth for the copy operation.
A similar correlation for the performance of the various configurations is obtained for the Xeon Phi, but the practical relevance is limited due to the low overall performance when using OpenCL.

%%%%%%%%%%%%%%%%%%%%%%%%%%%%%%% Intra- and Inter-Vendor %%%%%%%%%%%%%%%%%%%%%%%%%%%%%%%%%%%%%

\subsection{Intra- and Inter-Vendor Portability}

\begin{figure*}
 \centering
 \includegraphics[width=0.31\textwidth]{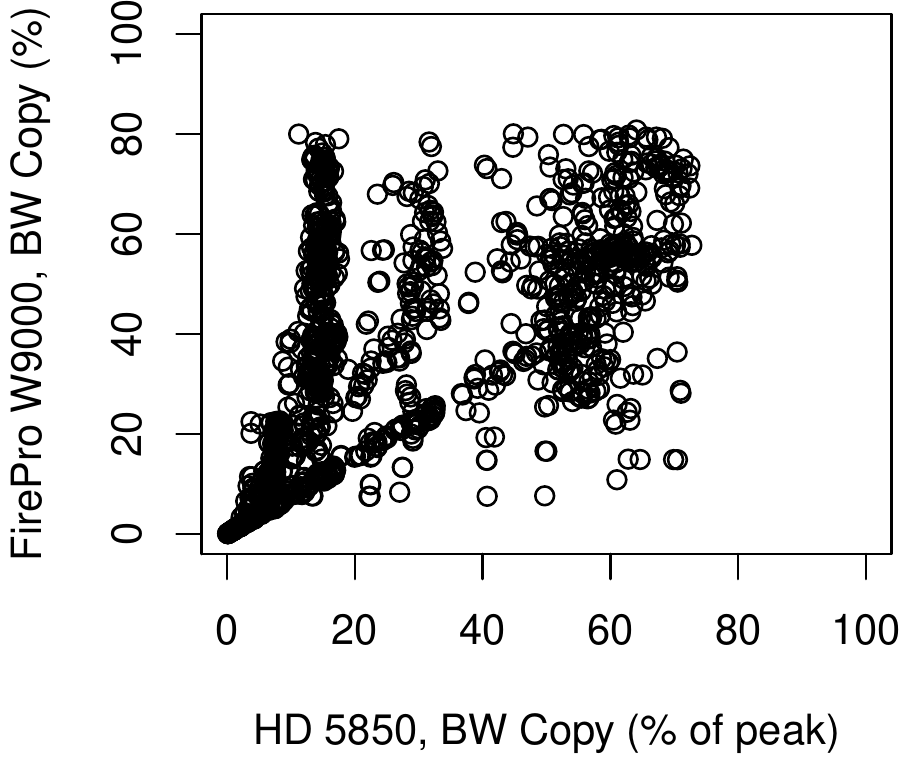} \hfill
 \includegraphics[width=0.31\textwidth]{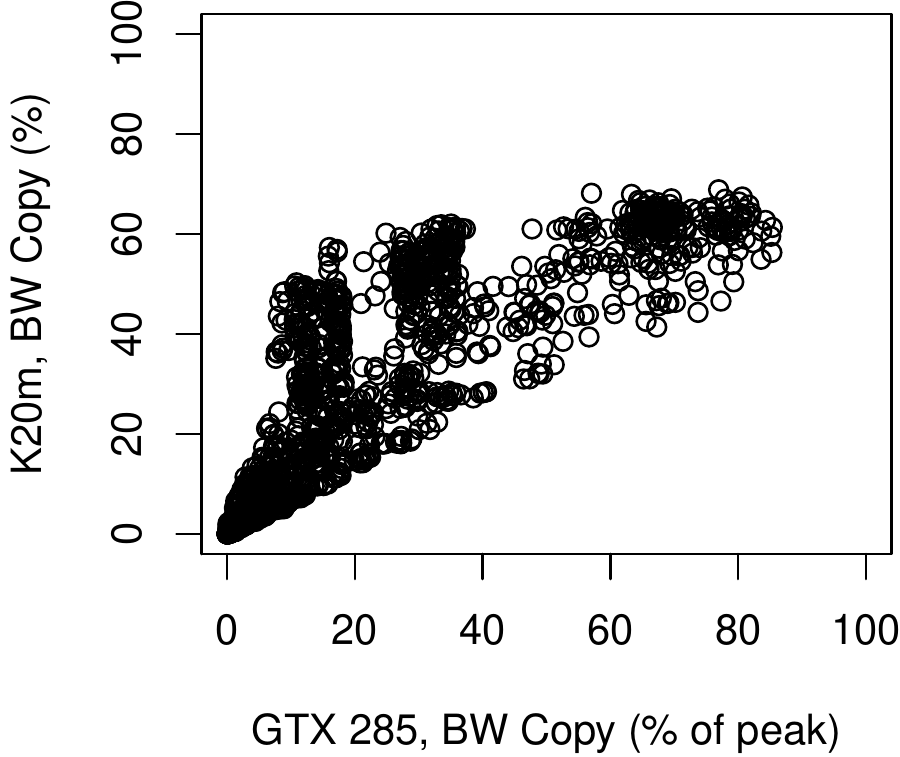} \hfill
 \includegraphics[width=0.31\textwidth]{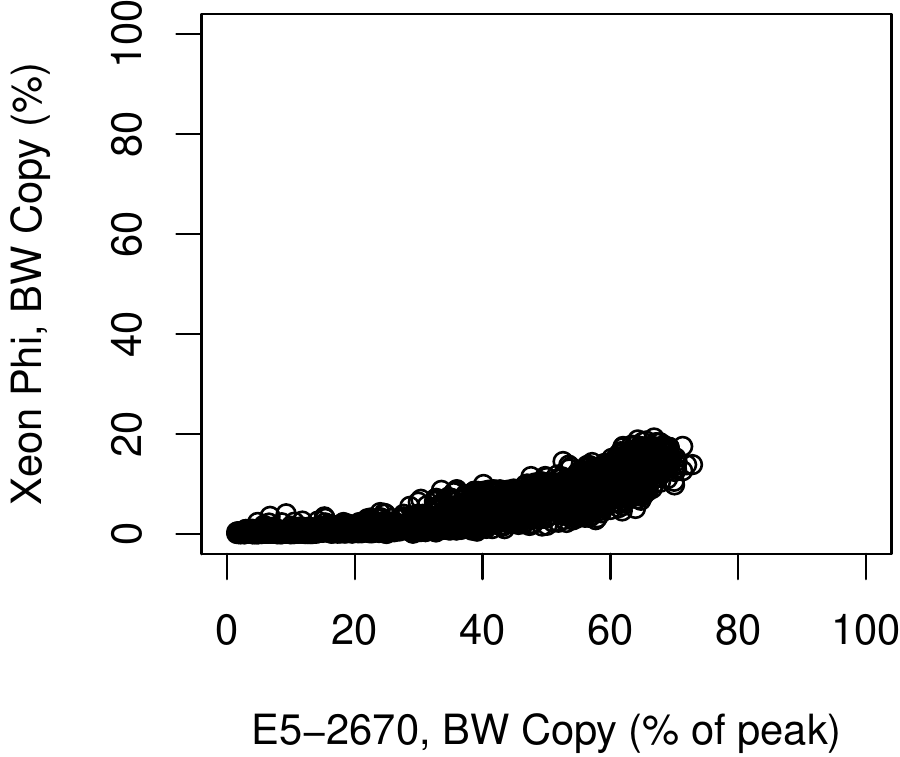} \\
 \caption{Comparison for the relative bandwidth (BW) of the copy kernel on hardware of the same vendor.
          A strong correlation is observed for devices from NVIDIA and INTEL, while the correlation on AMD hardware is weaker.
          In particular, good performance for the copy operation on one device has a high probability for good performance on the other device.}
 \label{fig:vendor-correlation}
\end{figure*}

\begin{figure*}
 \centering
 \subfigure{ \includegraphics[width=0.97\textwidth]{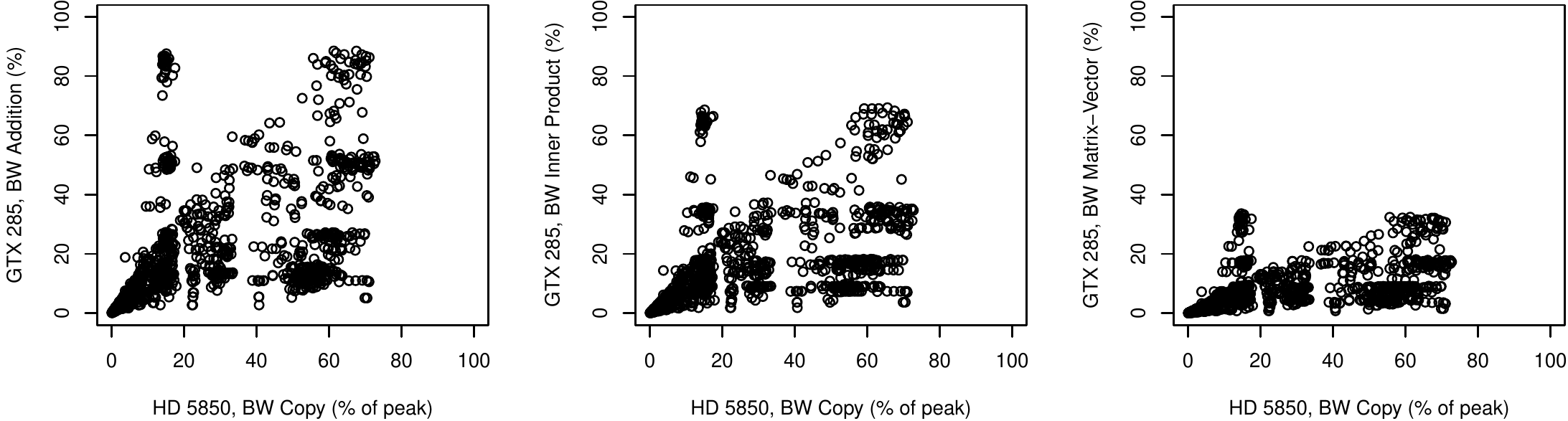} }
 \subfigure{ \includegraphics[width=0.97\textwidth]{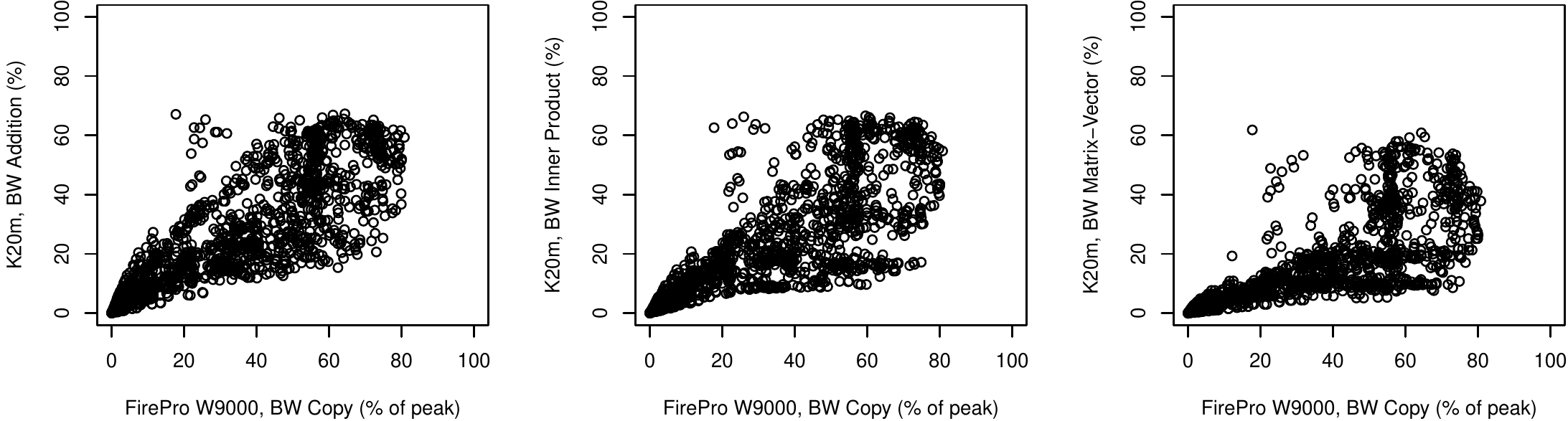} }
 \subfigure{ \includegraphics[width=0.97\textwidth]{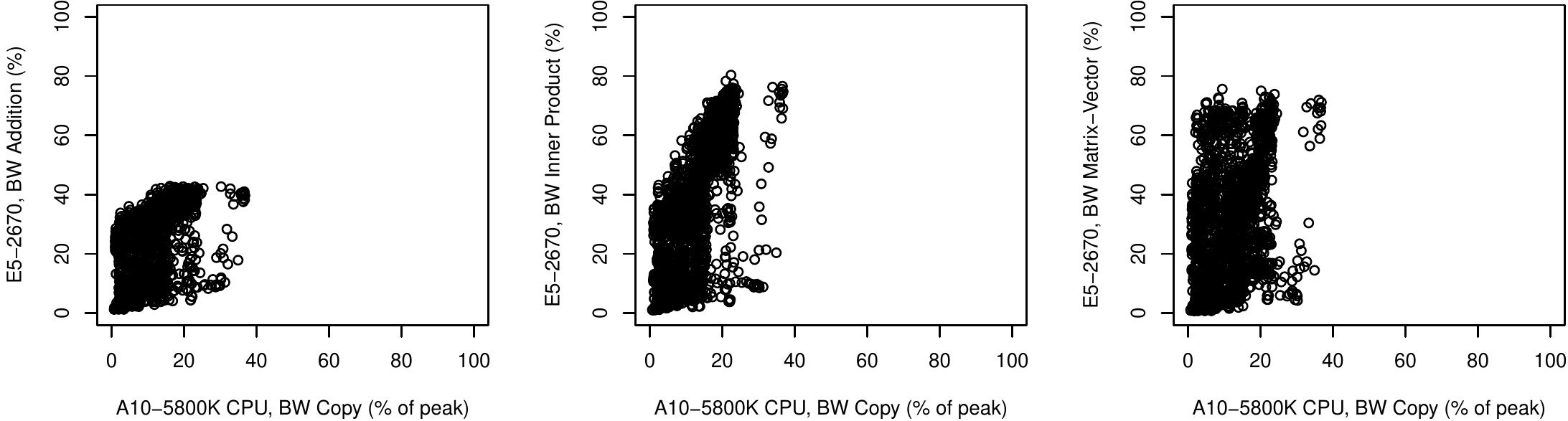} }
 \caption{Correlations of the relative bandwidth (BW) for kernel configurations of the copy kernel on GPUs and CPUs from AMD compared with the vector addition, inner product, and matrix-vector product kernels on NVIDIA GPUs and INTEL E5-2670 CPUs.
          Some of the fastest kernel configurations for the copy kernel also provide close-to-peak performance on the respective other device for all three operations. }
 \label{fig:cross-vendor-kernels}
\end{figure*}

After clear performance correlations among the operations considered on the same device have been identified, the next step is to study the performance correlation across different devices.
Figure~\ref{fig:vendor-correlation} compares the benchmark results for the copy kernel for devices of the same vendor.
The plot comparing the two AMD GPUs (HD 5850 vs.~W9000) not only shows that multiple configurations exist which are fast on both devices, but also shows many more configurations in the upper left half than in the lower right half.
This means that many configurations, which are slow on the older HD 5850, are significantly faster on the W9000, indicating that the newer hardware is more versatile.
A similar picture is obtained when comparing the two NVIDIA GPUs (GTX 285 vs.~K20m): The newer K20m often provides good performance for configurations which do not perform as well on the older GTX 285.
The comparison of the INTEL CPU and the Xeon Phi even shows that the best configurations for one device are also among the best on the other device and vice versa.

An overview of the best configuration obtained on each of the devices and compared on all other devices considered in this benchmark is given in Table~\ref{tab:best-configurations}.
The numbers demonstrate that it is not enough to tune on only one device, unless a performance penalty of at least $5-15\%$ on other devices is acceptable.
Moreover, on closer inspection one can identify two groups, within which better performance portability is obtained: 
One group consists of the CPUs (A10-5800K, E5-2670) and the Xeon Phi.
The second group contains the GPUs, within which better performance portability for hardware of the same vendor is observed.

While Figure~\ref{fig:vendor-correlation} only compares benchmark results for the copy operation across different devices, Figure~\ref{fig:cross-vendor-kernels} compares the benchmark results of the copy operation on one device with the addition, inner product, and matrix-vector product on another device.
About 40 configurations which are fast on both devices are found when comparing the HD 5850 and the GTX 285.
A comparison of the more recent W9000 and K20m exhibits a much more pronounced correlation: Most configurations show about the same performance on both devices.
However, the correlation is weaker when comparing the copy operation on the W9000 with the matrix-vector product on the K20m, yet a significant share of fast configurations for the copy operation on the W9000 is among the best configurations for the matrix-vector product on the K20m.
A comparison of the A10-5800K CPU with the E5-2690 CPU shows that most configurations perform significantly better on the E5-2670 than on the A10-5800K.
Again, it is possible to find configurations which are fast both for the copy operation on the A10-5800K and the addition, inner product, and matrix-vector product on the E5-2670.

\subsection{Best Configurations}
So far we have shown that among the best configurations for the copy kernel one can always find configurations with high performance for other operations.
As is shown in Table~\ref{tab:best-configurations-2}, one may even pick a single configuration for each device, which then achieves almost optimal performance for all four operations.
On all devices considered the performance is within $15$ percent of the peak bandwidth for the respective operation.
On NVIDIA and INTEL devices, the best configuration even results in performance within only five percent of the respective peak value.
We also note that several configurations with almost identical performance are obtained for each device.
For GPUs this also includes configurations without using any vector data types, which provide higher flexibility than vector data types.

% \begin{figure*}
%  \centering
%  \includegraphics[width=0.31\textwidth]{./figures/hd5850_double_gtx_285_double_copy.pdf} \hfill
%  \includegraphics[width=0.31\textwidth]{./figures/firepro_w9000_double_k20m_double_copy.pdf} \hfill
%  \includegraphics[width=0.31\textwidth]{./figures/a10_5800K_cpu_double_xeon_cpu_double_copy.pdf}
%  \caption{Correlations for the relative bandwidth (BW) of the copy kernel on hardware from different vendors.
%           The CPU-CPU comparison on the left as well as the GPU-GPU comparison on the right shows a clear correlation of bandwidths.
%           The mixed CPU-GPU comparison in the center shows no significant correlation.}
%  \label{fig:cross-vendor-copy}
% \end{figure*}

%%%%%%%%%%%%%%%%%%%%%%%%%%%%%%% Tables %%%%%%%%%%%%%%%%%%%%%%%%%%%%%%%%%%%%%

\begin{table*}
 \centering
 \begin{tabular}{|l||ccccccc|}
\hline
           & \multicolumn{7}{c|}{Best Kernel Configuration for} \\
                         & A10-5800K           & HD 5850             & W9000             & GTX 285            & Tesla K20m         & E5-2670           & Xeon Phi \\
\hline
\hline
AMD A10-5800K CPU         &   \textbf{36.8} \% & \hphantom{0}4.3 \% &             2.7 \% & \hphantom{0}2.3 \% &            12.7 \% &          18.0 \%  &            23.0 \% \\ %25.6
AMD Radeon HD 5850        & \hphantom{0}0.1 \% &   \textbf{72.7} \% &            64.1 \% &            61.3 \% &            55.5 \% &          14.1 \%  & \hphantom{0}4.2 \% \\ %128
AMD FirePro W9000         & \hphantom{0}4.4 \% &            73.1 \% &   \textbf{80.1} \% &            77.7 \% &            46.2 \% &          30.3 \%  & \hphantom{1}1.2 \% \\ %264
NVIDIA GeForce GTX 285    & \hphantom{0}2.8 \% &            67.9 \% &            73.3 \% &   \textbf{85.3} \% &            76.7 \% &          14.2 \%  & \hphantom{0}0.2 \% \\ %159
NVIDIA Tesla K20m         & \hphantom{0}9.5 \% &            47.1 \% &            50.5 \% &            61.3 \% &   \textbf{68.8} \% &          33.1 \%  & \hphantom{0}1.6 \% \\ %208
INTEL Xeon E5-2670 (dual) &            65.6 \% &            38.1 \% &            41.5 \% &            37.1 \% &            58.6 \% & \textbf{72.9} \%  &            66.8 \% \\ %102.4
INTEL Xeon Phi            &            14.1 \% & \hphantom{0}3.2 \% &             7.7 \% & \hphantom{0}4.6 \% & \hphantom{0}8.6 \% &          13.9 \%  &   \textbf{19.2} \% \\ %220
\hline
 \end{tabular}
 \caption{Memory bandwidth relative to the theoretical peak bandwidth for the vector copy operation.
          Performance penalties among the four GPUs (HD 5850, W9000, GTX 285, K20m) in the order of 5-15\% relative to peak are observed if the best configuration for one device is used on the other.
          The penalty is about the same for the two CPUs (A10-5800K, E5-2670) and the Xeon Phi.}
 \label{tab:best-configurations}
\end{table*}

\begin{table*}
 \centering
 \begin{tabular}{|l||ccccccc|}
\hline                 % 25.6           128                 264             159              208              102.4              220
                       & A10-5800K      & HD 5850         & W9000          & GTX 285        & Tesla K20m     & E5-2670           & Xeon Phi \\
\hline
\hline
Increment Type         & \emph{local}   & \emph{global}   & \emph{global}  & \emph{global}  & \emph{global}  & \emph{local}   & \emph{local} \\
Vector Length          & 2              & 8               & 4              & 1              & 2              & 4              & 16          \\
Local Work Size        & 1              & 128             & 64             & 128            & 256            & 1              & 1           \\
Workgroups             & 256            & 1024            & 160            & 80             & 1024           & 512            & 512         \\
BW Copy                & 36.7 (36.8)    & 59.5 (72.7)     & 73.9 (80.8)    & 85.3 (85.3)    & 66.8 (68.9)    & 69.5 (73.1)    & 18.8 (19.2) \\
BW Addition            & 45.2 (45.8)    & 61.5 (78.5)     & 77.5 (82.2)    & 88.5 (88.5)    & 67.1 (68.2)    & 40.1 (43.0)    & 20.3 (20.9) \\
BW Inner Product       & 60.3 (61.9)    & 84.8 (84.9)     & 82.7 (83.1)    & 69.0 (69.4)    & 62.6 (66.6)    & 80.6 (80.6)    & 13.5 (14.7) \\
BW Matrix-Vector       & 47.2 (57.3)    & 82.4 (82.4)     & 77.3 (80.6)    & 32.4 (33.6)    & 61.8 (61.8)    & 73.0 (75.9)    & 10.3 (12.0) \\
\hline
 \end{tabular}
 \caption{Summary of the best configurations in terms of average memory bandwidth for each device. 
          The memory bandwidth (BW) reported for each operations is given in percent and relative to the theoretical peak.
          Values in parentheses denote the best bandwidth among the 1900 configurations.
          The bandwidth obtained is in all cases within 15 percent of the achievable peak value, for NVIDIA and INTEL devices even within five percent.}
 \label{tab:best-configurations-2}
\end{table*}

\section{Summary \& Conclusion}

We investigated the performance of linear algebra operations on CPUs and GPUs of different vendors and found that kernel configurations, which work well for a simple vector copy operation, are also preferable for more complicated kernels.
Moreover, we have shown that it is sufficient to only consider kernel configurations within $15$ percent of the achievable peak performance for the copy operation in order to deduce fast configurations for more complicated operations such as inner products or even matrix-vector products.
This strategy can even be applied across hardware from different vendors.
Also, it has been observed that performance portability is better on newer hardware, which indicates that initial performance portability issues with OpenCL were also caused by immature hardware - and possibly software and drivers.

In summary, our study provides the following guidelines for OpenCL developers who aim to run their OpenCL-enabled software on a broad range of consumer hardware:
For best performance, CPUs and GPUs should be considered separately.
On CPUs, each work item should operate on large chunks of consecutive data and either very small or very large workgroup sizes should be considered.
On GPUs, workgroup sizes of $128$ and $256$ are preferred.
A larger number of workgroups is advisable on NVIDIA GPUs, whereas a number of about $128$ already shows good performance on AMD GPUs.
Vector data types for operations limited by memory bandwidth are not required on GPUs, but may provide mild benefits for CPUs.
If, however, a performance reduction of about ten percent is acceptable, one may also use the same kernel structure for both CPUs and GPUs, reducing the overall implementation effort.
Our benchmark results suggest that the best trade-off between performance and simplicity for memory bandwidth limited operations is to let each work item operate on consecutive data, to use 128 or 256 work items per workgroup, to execute the kernel with least 128 workgroups, and to ignore vector data types.

%Instead of relying on brute-force autotuning, the strategy derived in our study allows for a considerable reduction of the search space, reducing the effort for portable performance substantially and identifying sweet spots.

\section*{Acknowledgments}

K.~Rupp, T.~Grasser, and A.~J\"ungel gratefully acknowledge support from the Austrian Science Fund (FWF), grant P23598.
F.~Rudolf and J.~Weinbub have been supported by the European Research Council (ERC) through the grant \#~247056 MOSILSPIN.
The authors thank Joachim Sch\"oberl (TU Wien) for providing access to a machine equipped with NVIDIA Tesla K20m GPUs.
Moreover, the authors are grateful to AMD for providing a FirePro W9000.

\bibliographystyle{plain}
\bibliography{ref.bib}

% \begin{thebibliography}{9}
% 
% \bibitem{OpenCL}
% OpenCL. \url{http://www.khronos.org/opencl/}
% 
% \bibitem{Tomov2012}
% J.~Kurzak \textit{et al.}: \textit{IEEE Par. Distr. Systems}, 2012.
% 
% \bibitem{Matsumoto2012}
% K.~Matsumoto \textit{et al.}: \textit{MCSoC-12}, 2012.
% 
% \bibitem{Atlas}
% ATLAS library. \url{http://math-atlas.sourceforge.net/}
% 
% \bibitem{Tillet:Performance-portable-linear-algebra}
% Ph.~Tillet \textit{et al.}: \textit{Proc.~HotPar'13}, 2013.
% 
% \bibitem{ViennaCL}
% ViennaCL library. \url{http://viennacl.sourceforge.net/}
% 
% %\bibitem{Spiral}
% %M.~P\"uschel \textit{et al.}: \textit{Int. J. High Perform. Comput. Appl.}, 2004.
% 
% \bibitem{Bell:spmv}
% N.~Bell, M.~Garland: \textit{Proc. Supercomputing}, 2009.
% 
% \end{thebibliography}

\end{document}